\documentclass[11pt]{article}
\usepackage{graphicx}
\usepackage{amsmath,amssymb}
\usepackage{todonotes}

\newcommand{\coo}{\mbox{CO${}_2$}}
\newcommand{\R}{\mathbb{R}}
\newcommand{\Prob}{\mathbb{P}}
\newcommand{\calA}{\mathcal{A}}
\newcommand{\ppm}{\mathit{ppm}}

\usepackage{fullpage}

\begin{document}

\title{Extreme sensitivity and climate tipping points}

\author{Peter Ashwin \and Anna S. von der Heydt}




\author{Peter Ashwin\thanks{P.Ashwin@exeter.ac.uk}\\
Centre for Systems, Dynamics and Control,\\
Department of Mathematics, University of Exeter,\\
Exeter, EX4 4QF UK   
%
\and
A. S. von der Heydt\thanks{A.S.vonderHeydt@uu.nl}\\
Institute for Marine and Atmospheric Research,\\
Department of Physics, also Centre for Complex Systems Studies, \\
Utrecht University, Princetonplein, \\
Utrecht, The Netherlands
}

\date{\today}

\maketitle

\begin{abstract}
A climate state close to a tipping point will have a degenerate linear response to perturbations, which can be associated with extreme values of the equilibrium climate sensitivity (ECS). In this paper we contrast linearized (`instantaneous') with fully nonlinear geometric (`two-point') notions of ECS, in both presence and absence of tipping points. For a stochastic energy balance model of the global mean surface temperature with two stable regimes, we confirm that tipping events cause the appearance of extremes in both notions of ECS. Moreover, multiple regimes with different mean sensitivities are visible in the two-point ECS. We confirm some of our findings in a physics-based multi-box model of the climate system.
\end{abstract}



\section{Introduction}

The equilibrium climate sensitivity (ECS) is widely used as a measure for expected future global warming. Following Charney's definition \cite{Charney:1979}, the ECS is the increase in global mean surface temperature (GMST) per radiative forcing change after the fast-acting feedback processes in the Earth System reach equilibrium. Fast-acting means here that those processes are faster than the time-horizon for global mean temperature evolution that interests us, typically taken to be 100 years \cite{rohling:2012}. 

The value of ECS remains not very well constrained, as the expected warming per doubling of atmospheric \coo{}{} still contains a considerable uncertainty of 1.5 -- 4.5$^{\circ}$C \cite{IPCC2013}. In fact, this range has not changed much since first ECS estimates based on energy balance arguments, despite enormous developments in climate modelling, and improved observational methods \cite{Knutti:2008}. In particular, large temperature changes and dangerous climate change as a consequence of increased atmospheric \coo{} cannot be excluded. For example, climate observations from the instrumental period have not narrowed down the range of expected climate change mainly because of uncertainties in quantification of the forcing \cite{schwartz:2012}. Recently there have been concerted attempts to constrain ECS using emergent constraints in climate models \cite{Cox:2018}.

Some sources of uncertainty for ECS lie in the classical measurement or model uncertainty, although in particular for observations the quantification of the applied forcing generally contains the largest uncertainties. Furthermore, by going back in time further than the instrumental period (e.g. last millennium, glacial cycles or even millions of years of palaeoclimate data) the uncertainty in both the forcing and the global mean temperature response becomes significant \cite{Knutti:2008}. 
There has been a debate as to the cause of the uncertainty and extremes of the ECS distribution, in particular the long tail towards high sensitivity values. On the one hand, \cite{RoeBaker:2007} suggest it is an inevitable consequence of nonlinear transformation of normally distributed feedbacks that appear in the denominator when calculating ECS. On the other hand, \cite{zaliapin2010another} suggest they are a sign of `tipping points' owing to nonlinearities in the system - this has generated a lively debate. In this paper we highlight (a) the notion of ECS can usefully be generalised to a truly nonlinear geometric notion: the two-point sensitivity and (b) the distribution of ECS values is a valuable tool for characterising both state-dependent (feedback) dynamics and tipping of the climate system.

The climate system exhibits both internal and forced variability on many timescales. The consequence is that any `equilibrium' is only relative to fixing part of the feedback processes that are internal to the climate system, in particular the `slower' part. This requires an assumption that a time scale separation (into fast and slow processes) exists and the time scale of interest sits between fast and slow. For climate model simulations and observations of the last century there might be a time horizon where this is a reasonable assumption \cite{schwartz:2012}, but as we include palaeoclimate data and model simulations into the estimate of ECS this assumption needs to be carefully evaluated. In particular, methods to estimate ECS from palaeoclimate data or models differ from those of (short) climate model simulations; the latter generally derive ECS from the decay of the energy imbalance at the top of the atmosphere induced by a instantaneous doubling or quadrupling of \coo{} \cite{Gregory:2004io}; palaeoclimate reconstructions instead make the assumption that the reconstructed climate is in (energetic, short time scale) equilibrium and compare different of these `equilibria' to each other for estimating ECS. Without compensating for slow feedback processes, palaeoclimate records give the so-called Earth System Sensitivity (ESS) that includes the effect of slow processes and boundary conditions (e.g. geography, vegetation and land ice) \cite{Lunt2010ng}. If estimates of these slow processes are available then ECS can be estimated from the ESS under an assumption of time scale separation \cite{rohling:2012,Heydt:2017}.
 
 Note that the ECS is usually thought of as a linearized response of the GMST to perturbations in the radiative balance of the earth. Next to incoming (short-wave) and outgoing (long-wave) radiation, feedback processes in the climate system play an important role in determining the ECS. In their sum these feedback processes tend to enhance ECS (net positive feedback) and associated time scales vary from fast to very slow. Examples of fast feedback processes include cloud feedbacks, water vapour feedback and sea-ice processes. The strength of each of these feedback processes depends on the background (long-term mean) climate state \cite{Heydt2014} and it is therefore not surprising that the sum of the fast feedback processes varies over time in particular when considering climate states far back in time and under very different boundary conditions. For example, from palaeoclimate records together with ice sheet modelling, it has been found that ECS varies considerably between glacial and interglacial states \cite{Koehler:2017jj,Friedrich:2016dn,Koehler:2015}. Both present-day and Palaeogene climate model simulations suggest state-dependence of ECS due to feedback processes \cite{Caballero2013,pfister:2017iz}. For example, Transient Climate Response (TCR) considers the deep ocean warming as a slow process. 

Abrupt climate shifts have occurred in the past climate system and therefore seem likely to occur in the future for a variety of reasons \cite{Lenton:2008de,Steffen:2018}. In the present climate system, potential tipping elements have been identified some of which may have a considerable impact on future values of GMST \cite{Lenton:2008de,Drijfhout:2015hj}. Even those tipping elements that have little affect on the GMST may cause significant regional damage and/or contribute to global mean climate change by triggering cascades of transitions involving other tipping elements \cite{Dekker:2018bs}. Across such an abrupt transition there is a breakdown of the assumption of a linear response to perturbations, suggesting that the ECS does not adequately represent the temperature response to radiative perturbations \cite{Heydt2016rev,BlochJohnson:2015hm}. In practice, when deriving ECS from palaeoclimate time series, which include abrupt transitions, these shifts may lead to extreme values of ECS.

In this paper we show that more general notions of ECS can be useful in understanding the response of a climate state to changes in radiative forcing - in addition to an `instantaneous' linearised notion of ECS we explore `incremental' and `two-point' climate sensitivities that are distributions related to dynamic properties of the climate system: they characterise the geometry of the dynamics and are not simply estimators of a `mean ECS'.  In fact the distributions of sensitivities reflect the intrinsic uncertainty due to climate system dynamics. The paper is organized as follows: Section~\ref{sec:sens} introduces these notions of ECS and relates them to the underlying climate dynamics. We illustrate these concepts using a global energy balance model in Section~\ref{sec:EBM}. In particular, we relate properties of extremes of the ECS to the presence of tipping points and multistability.  Section~\ref{sec:discuss} finishes with a discussion of conclusions and some challenges for the future. Appendix~\ref{sec:GT} extends results of \cite{Heydt:2017} and examines extremes of this sensitivity associated with tipping points in a more realistic physics-based multi-box model of the glacial cycles by Gildor and Tziperman \cite{Gildor:2001}. 

\section{Sensitivities and the climate attractor}
\label{sec:sens}

In order to understand variability, abrupt transitions and response to perturbations we consider the climate system as a high-dimensional multiscale complex dynamical system whose evolving trajectories form a {\em climate attractor}. The ECS can be defined on this attractor and regimes or states may be identified where a linear approximation of the response may be reasonable. Tipping points visible in the GMST will show up as large but occasional shifts between different `climate regimes' of the attractor, or indeed different attractors. We visualise the attractor by projection onto climate observables relevant for determining ECS, i.e. the GMST $T$ and the radiative forcing $R$ per unit area \cite{Heydt:2017}. Consider the energy balance model 
\begin{equation}
c_T \frac{d T}{d t} = R_{\mathrm{forcing}}  + R_{\mathrm{slow}} + R_{\mathrm{fast}} - R_{\mathrm{OLW}},
\label{e:energybalance}
\end{equation}
where the left hand side represents the rate of change of the global mean surface temperature $T$ (with specific heat capacity $c_T$) and on the right hand side $R_{\mathrm{forcing}}$ is the (external) radiative forcing (including changes in \coo{}), $R_{\mathrm{slow}}$ ($R_{\mathrm{fast}}$) is the radiative perturbation due to all slow (fast) feedback processes within the climate system and $R_{\mathrm{OLW}}$ is the outgoing longwave radiation, respectively. 
Following the formalism of \cite{rohling:2012}, the specific climate sensitivity is
\begin{equation}
S_{\mathrm{forcing},\mathrm{slow}} =  \frac{\Delta T}{\Delta R_{\mathrm{forcing}}+\Delta R_{\mathrm{slow}}} \approx \frac{dT}{d(R_{\mathrm{forcing}}+R_{\mathrm{slow}})},
\label{eq:Sspecific}
\end{equation}
which equals the Charney sensitivity $S$ if $\Delta R_{\mathrm{slow}}$ is the sum of all slow feedback processes contributing to the ECS (and under the assumption of time scale separation). In practise, only some of the slow processes are accessible from palaeoclimate records (e.g. only land ice), in which case the specific climate sensitivity is only an approximation of the Charney sensitivity \cite{rohling:2012} (e.g. $S_{[CO_2,LI]}$ is the specific climate sensitivity considering only land ice changes as slow feedback). This ECS gives a linear prediction for change in temperature:
\begin{equation}
T'=T+ S_{\mathrm{forcing},\mathrm{slow}}\, \left(\Delta R_{\mathrm{forcing}}+\Delta R_{\mathrm{slow}}\right).
\label{eq:linpred}
\end{equation}
For a specific energy balance model including regime shifts we can explicitly calculate ECS for the different regimes, see section \ref{sec:EBM}. We note that several other authors have highlighted the need to improved notions of ECS: this includes \cite{Chekroun:2011} who propose to use a measure-based approach to understand climate sensitivity and  \cite{dijkstra2015sensitivity} who consider conditional climate sensitivities constrained by temperature, coupled with resilience measures for switching to other regimes.

\subsection{Observation of the climate attractor}

We consider the climate system as a high dimensional dynamical system that evolves along trajectories $x(t)$ according to a smooth flow
\begin{equation}
x(t)=\varphi_t(x_o)
\label{e:climatesystem}
\end{equation}
where $x\in X$ represents the instantaneous state of the climate system in some high dimensional state space and $\varphi_t(x_0)$ evolves the initial state $x_0$ along by a time $t$. The global mean temperature $T:X\rightarrow \R$ and radiative forcing $R:X\rightarrow \R$ are considered to be observables of the underlying dynamical system on $X$. We assume that the dynamics of $x$ are stationary, i.e. that there is a natural probability measure $M$ on $X$ such that typical trajectories $x(t)$ of (\ref{e:climatesystem}) satisfy \begin{equation}
\lim_{t\rightarrow \infty} \frac{1}{t}\int_{s=0}^{t} p(x(s)) \, ds =\int_{x} p(x) \, dM(x).
\label{e:average}
\end{equation}
for typical $x_0$ and any integrable observable $p:X\rightarrow \R$, i.e. the long-time average of $p$ can be computed using an ergodic hypothesis, by averaging over the measure $M$ in phase space. This implies that, for any open set $\calA\subset X$, the long-term average proportion of time a typical trajectory spends in $\calA$ is $M(\calA)$. For small enough perturbations, linear response theory suggests a linear change in mean observables: see for example \cite{lucarini2018}.

In \cite{Heydt:2017} it is supposed there is a stationary measure $\mu$ of points in the $(\Delta R_{[CO_2,LI]},T)$-plane according to how often they are visited over asymptotically long times, i.e. for any measurable subset $\calA\subset \R^2$ we define
\begin{equation}
\mu(\calA):=\lim_{t\rightarrow \infty} \frac{1}{t}\int_{s=0}^{t} \chi_{\calA}(\Delta R_{[CO_2,LI]}(s),T(s))\,ds.
\label{eq:freqTvsR}
\end{equation}
for typical initial condition, where $\chi_{\calA}$ is the indicator function, $\chi_{\calA}(\Delta R,T)=1$ if $(\Delta R,T)\in \calA$ and $=0$ otherwise. Note that applying (\ref{e:average}) with $p(x)=\chi_{\calA}(x)$ (where $\chi_{\calA}(x)=1$ if $x\in {\calA}$ and $0$ otherwise) gives
\begin{equation}
\mu({\calA})=M(\{x~:~( R_{[CO_2,LI]}(x),T(x))\in {\calA}\}).
\label{e:projection}
\end{equation}
In other words, the measure $\mu$ is simply a projection of a natural measure $M$ on the `climate attractor' onto the two observables $( R_{[CO_2,LI]},T)$. In general we will consider throughout
$$
\Delta R_{[CO_2,LI]}=R_{[CO_2,LI]}-\tilde{R}_{[CO_2,LI]}
$$
i.e. the change in radiative forcing  relative to some fixed 
reference level $\tilde R_{[CO_2,LI]}$ usually chosen as the level during the pre-industrial climate.

 \subsection{Incremental and two-point sensitivities} 

In order to predict the temperature at some fixed future time-horizon $\Delta t$ in response to a change in radiative forcing, we consider the quotient (\ref{eq:Sspecific}) for fixed changes in time. In this case we can view the distribution of what we call {\em incremental sensitivities} as the spread of trajectories from the current estimated values of instantaneous $(\Delta R,T)$, where from now on we write the radiative forcing corrected for slow feedback as $R$. On the other hand we can consider a time-independent choice of pairs of points on the climate attractor to obtain {\em two-point sensitivities}.

\begin{figure}[!ht]
\centering
\includegraphics[width=6cm]{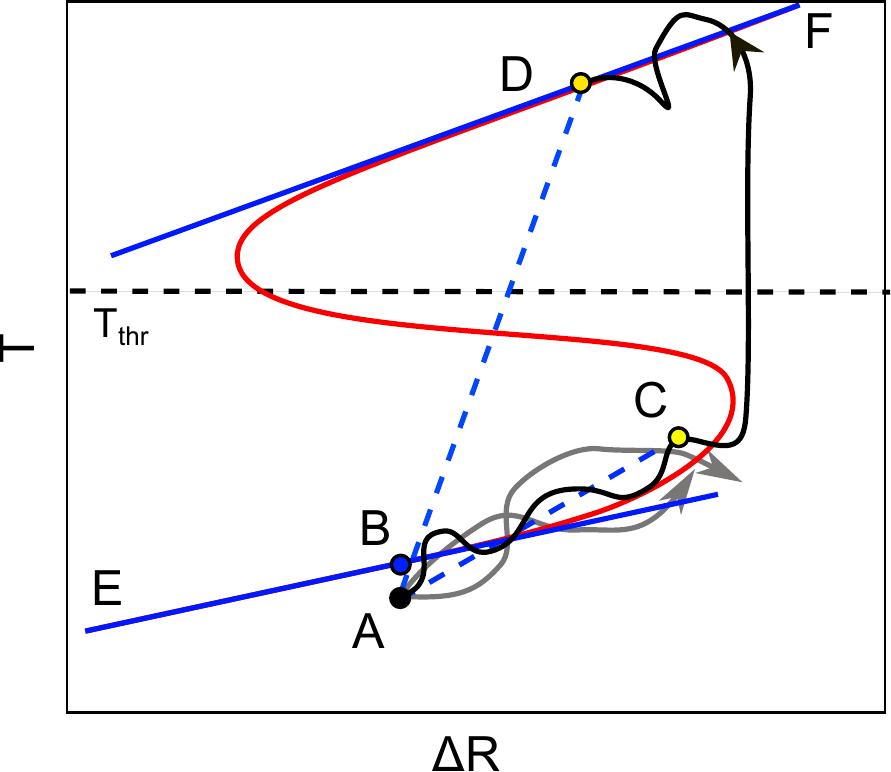}

\caption{Schematic diagram to demonstrate the instantaneous, incremental and two-point sensitivities for an ensemble of trajectories starting at point $A$ that evolves on the climate attractor. The equilibrium of a mean energy balance model is shown as a red curve. For a specific trajectory (shown in black) in the ensemble the slopes of $AC$ and $AD$ correspond to incremental sensitivities (for fixed $\Delta t$) or two-point sensitivities (for varying $\Delta t$). The instantaneous sensitivity is the slope of the tangent to the closest solution in the equilibrium model at $B$. Note that the cold regime $T\leq T_{\mathrm{thr}}$ and the warm regime $T>T_{\mathrm{thr}}$ have different asymptotic  (instantaneous) sensitivities corresponding to slopes $EB$ and $DF$ respectively.}
\label{fig:schem}
\end{figure}

Let us assume the current state at time $t=0$ of the climate system is given by a measure $\sigma_0$ on some high dimensional phase space $X$ (that projects onto a point A in Fig.~\ref{fig:schem}). Note that this will always be a measure rather than a point because of lack of knowledge of sub-grid parametrized processes (e.g. \cite{Tantet2018}) but it will project onto the current values $(\Delta R_0,T_0)=(\Delta R(x),T(x))$ for all $x$ in the support of $\sigma_0$. As time progresses, this state will spread to give a measure at time $t$ that is
$$
\sigma_t({\calA})= \sigma(\varphi_{-t}({\calA}))
$$
for any ${\calA}\subset X$ (Fig.~\ref{fig:schem} shows a trajectory in black and others from the ensemble starting at $A$ in grey). The {\em incremental sensitivity} for a time interval $\Delta t$ is then 
\begin{equation}
S^{\Delta t}_0(x)=\frac{T(\varphi_{\Delta t}(x))-T_0}{\Delta R(\varphi_{\Delta t}(x))-\Delta R_0}
\label{eq:Sinc}
\end{equation}
with distribution
$$
\Prob(S_0^{\Delta t} \in {\calA})= \sigma(\{x~:~S^{\Delta t}_0(x)\in {\calA}\}).
$$
Over long time, if there is decay of correlations and mixing of trajectories on the climate attractor \cite{Tantet2015,Tantet2018, lucarini2018} then $\sigma_{\Delta t}\rightarrow M$ in the weak sense as $\Delta t\rightarrow \infty$, and so we expect the distribution of long-term incremental sensitivities for $\Delta t\rightarrow \infty$ become time-independent for typical trajectories within the attractor:
\begin{equation}
\Prob(S_0^{\infty} \in {\calA})= M(\{x~:~S^{\infty}_0(x)\in {\calA}\}).
\label{eq:S0inf}
\end{equation}
where
$$
S_0^{\infty}(x)=\frac{T(x)-T_0}{\Delta R(x)-\Delta R_0}.
$$
Note that (\ref{e:projection}) means that the distribution of long-term sensitivities starting at $(\Delta R_0,T_0)$ can be written in terms of the geometry of the projected measure $\mu$
\begin{equation}
\Prob(S_0^{\infty} \in {\calA})= \mu(\{(\Delta  R_1,T_1)~:~S^{\infty}_{0,1}\in {\calA}\})
\label{e:longtermsens}
\end{equation}
where we define the {\em two-point sensitivity} as
\begin{equation}
S^{\infty}_{0,1}=\frac{T_1-T_0}{\Delta R_1-\Delta R_0}.
\label{eq:S01}
\end{equation}

The distribution of long-term incremental sensitivities (\ref{e:longtermsens}) for a generic choice of the initial climate state suggests \cite{Heydt:2017} a time-independent notion of climate sensitivity that can be found by picking pairs of points $(R_{0,1},T_{0,1})$ independently distributed according to $\mu$ and evaluating (\ref{eq:Sspecific}). 

This means that for any ${\calA}\subset \R$ we can use $\mu$ to assign a probability to the sensitivity being in ${\calA}$:
\begin{equation}
\Prob(S^{\infty}_{0,1}\in {\calA}):=\mu\times\mu\left(\left\{ (\Delta R_0,T_0),(\Delta R_1,T_1)~:~S^{\infty}_{0,1} \in {\calA}\right\}\right).
\label{eq:Sdist}
\end{equation}
with $S^{\infty}_{0,1}$ defined as in (\ref{eq:S01}). This gives, in some sense, a maximal set of possibilities for the sensitivities in that it compares the observables $T$ and $\Delta R$ over all possible time points and possible trajectories of the system. This is comparable to the conditional climate sensitivity of \cite{dijkstra2015sensitivity} except rather than dividing into regimes, they restrict to deviations of temperature at most $\delta_T$ from $T_0$.  In the case that the sensitivity is fixed at $S_0$, note that $S^{\infty}_{0,1}$ is a Dirac $\delta$-distribution centred at $S_0$.

\subsection{Sensitivities and climate regimes}

By partitioning the climate attractor into a number of regimes, we can condition the sensitivities on staying within a regime, or undergoing a transition between regimes. By making an optimal partition of the attractor projected into $(\Delta R,T)$ space we can hope to find localised distributions of sensitivities for pairs in the same regime. As in \cite{Heydt:2017} we consider these sensitivities conditional on climate regime by partitioning $\mu$ into two distributions
$$
\mu=\mu_C+\mu_W
$$
corresponding to being in a cold (C) or warm (W) state. In our case we set
\begin{align*}
\mu_{C}({\calA})&=\mu({\calA}\cap \{(\Delta R,T)~:~T\leq T_{thr}\}),\\
\mu_{W}({\calA})&=\mu({\calA}\cap \{(\Delta R,T)~:~T>T_{thr}\})
\end{align*}
for any measurable ${\calA}\subset \R^2$ and some threshold temperature $T_{thr}$. As in \cite{Heydt:2017} we define distributions of conditional sensitivities by
\begin{equation}
\Prob(S^{WW}\in {\calA}):=\mu_W\times\mu_W\left(\left\{ (\Delta R_0,T_0),(\Delta R_1,T_1)~:~S^{\infty}_{0,1} \in {\calA}\right\}\right)
\label{eq:SWW}
\end{equation}
Conditional sensitivities for changes of regime are for example
\begin{equation}
\Prob(S^{CW}\in {\calA}):=\mu_C\times\mu_W\left(\left\{ (\Delta R_0,T_0),(\Delta R_1,T_1)~:~S^{\infty}_{0,1} \in {\calA}\right\}\right)
\label{eq:SCW}
\end{equation}
The distribution of sensitivities (\ref{eq:Sdist}) is then a sum of the four conditional sensitivities 
$$
\Prob(S^{\infty}_{0,1}\in {\calA})=
\Prob(S^{CC}\in {\calA})+
\Prob(S^{CW}\in {\calA})+
\Prob(S^{WC}\in {\calA})+
\Prob(S^{WW}\in {\calA}).
$$
Moreover, (\ref{eq:SCW}) means that we have a symmetry
$$
\Prob(S^{CW}\in {\calA})=\Prob(S^{WC}\in {\calA}).
$$
The distributions of $S^{CW}$ and $S^{WC}$ correspond to choices of pairs across the two regimes: these distributions are associated with `tipping between regimes'. Even though the two-point sensitivities may measure states very far apart in time, we find that extreme values of the sensitivity are usually associated with choice of points from two different regimes. 

\section{Sensitivity and tipping in climate models}
\label{sec:EBM}

To illustrate the notions of instantaneous and two-point sensitivities, we consider a conceptual energy balance model: a more complex model is briefly discussed in Appendix~\ref{sec:GT}. We consider a variant of the Budyko-Ghil-Sellers energy balance model \cite{budyko1969effect,ghil1976climate,sellers1969global} for GMST. This model builds on \cite{dijkstra2015sensitivity,zaliapin2010another} and has multiple regimes with state-dependent sensitivity in each. It is a special case of (\ref{e:energybalance}) for global mean surface temperature $T(t)$ with atmospheric CO$_2$ concentration $C(t)$ as a parameter:
\begin{equation}
\label{eq:EBMdet}
c_T\, \frac{dT}{dt}=  F(T,C):=\left[Q_0(1-\alpha(T))+A\ln\left(\frac{C}{C_0}\right)-\epsilon(T)\sigma T^4\right]
\end{equation}
For this equation, $Q_0$ represents the solar input modulated by the temperature-dependent albedo $\alpha(T)$. The change in radiative forcing due to atmospheric greenhouse gases is
$$
\Delta R_{[CO_2]}=A\ln(C/C_0).
$$
where $A=5.35$~Wm$^{-2}$ is the direct forcing effect of \coo{} and $C_0$ represents pre-industrial \coo{} levels. Finally, the outgoing long wave radiation $\sigma T^4$ is modified by a temperature-dependent emissivity $0<\epsilon(T)< 1$. 

\begin{figure}
    \centering
    \includegraphics[width=11cm]{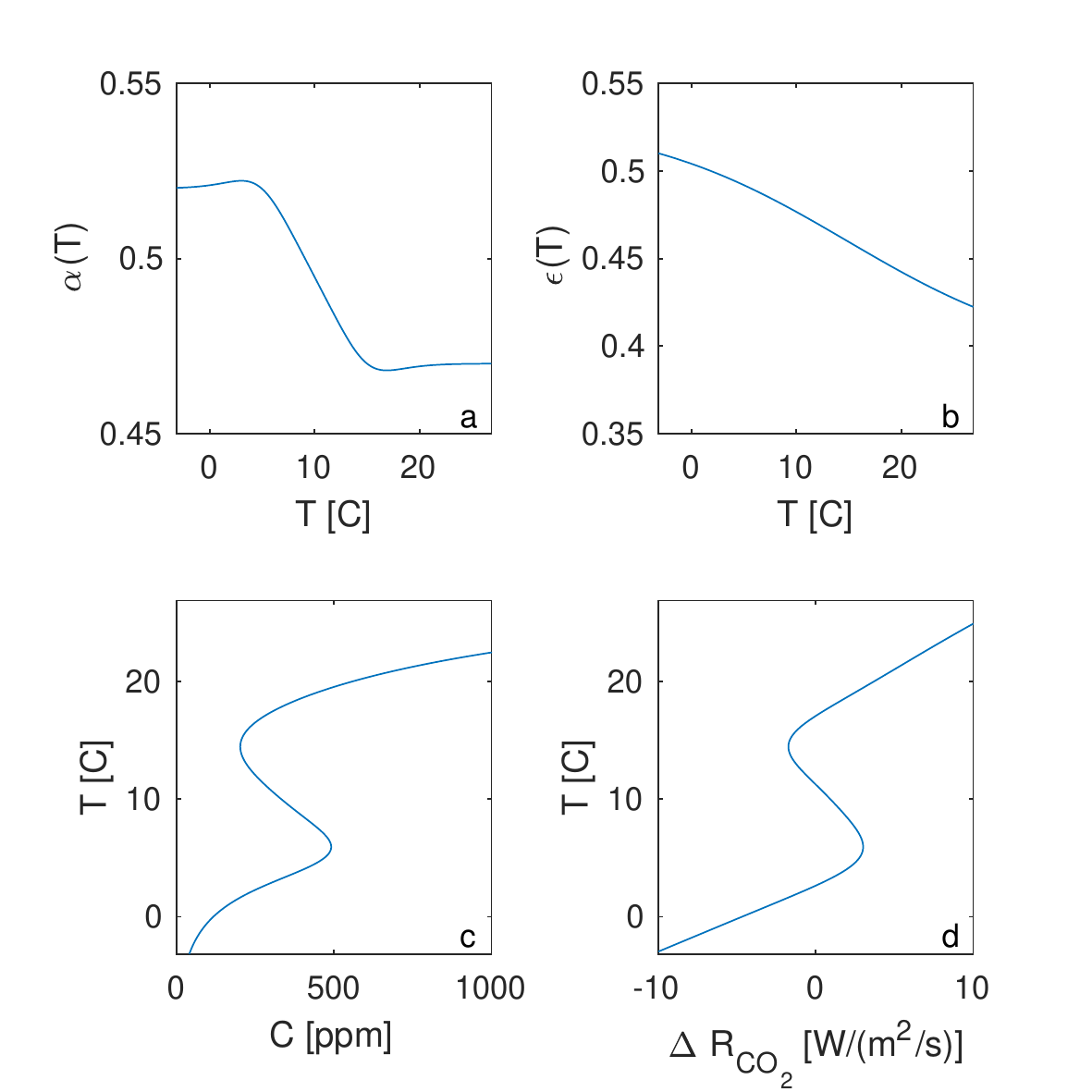}
    \caption{Behaviour of the equilibrium energy balance model (\ref{eq:EBMdet}) with parameters as in Table~\ref{tab:EBMparams}. (a) temperature-dependence of albedo $\alpha(T)$ and (b) emissivity $\epsilon(T)$; (c) \coo{} and (d) radiative forcing levels necessary to give temperature equilibria, corresponding to (\ref{eq:EBMequil}) and (\ref{eq:EBMDRTeqm}), respectively. Note the region of multistability, and temperature-dependence of the sensitivity corresponding to slopes in the bottom right figure.
    \label{fig:alphaeps}}
\end{figure}

We consider a temperature-dependent emissivity  decreasing from one plateau to a lower one because of changes in water vapour and cloud feedbacks. There are other choices \cite{zaliapin2010another}, but for simplicity we assume here
\begin{equation*}
\epsilon(T)= \epsilon_1+\frac{\epsilon_2-\epsilon_1}{2}\left[1+\tanh\left(\frac{T-T_0}{T_{\epsilon}}\right)\right].
\end{equation*}
where $\epsilon_1$ and $\epsilon_2$ are the limit emissivities for low and high temperatures respectively, $T_0$ is the threshold and $T_{\epsilon}>0$ corresponds to the range of temperatures over which there is variation (see Figure~\ref{fig:alphaeps}b). 

Note that water vapour feedback is sometimes included in the \coo{} term, resulting in an additional constant and modified $A$ \cite{hogg2008glacial,dijkstra2015sensitivity}. Here we separate radiative forcing due to \coo{} and temperature-dependent water feedbacks in emissivity. As in \cite{dijkstra2015sensitivity}, we assume that the albedo varies with temperature due to changes in land-ice feedback processes: we assume there are threshold temperatures $T_1<T_2$ associated with changes of albedo $\alpha (T)$ and define a function
\begin{equation}
\Sigma(T)= \frac{(T-T_1)}{T_2-T_1}H(T-T_1)H(T_2-T)+H(T-T_2)
\end{equation}
that switches from $0$ for $T<T_1$ to $1$ for $T>T_2$: $H(T)$ is approximately a Heaviside unit step function and we use a smooth approximation
$$
H(T)=(1+\tanh(T/T_{\alpha}))/2
$$
as in \cite{dijkstra2015sensitivity}. As in that paper, we write the albedo
\begin{equation*}
\alpha(T) = \alpha_1(1-\Sigma(T))+\alpha_2\Sigma(T)
\end{equation*}
so that it changes smoothly from a higher albedo $\alpha_1$ in the presence of more ice surface ($T<T_1$) to a lower $\alpha_2$ in the presence of more ocean surface ($T>T_2$), see Figure~\ref{fig:alphaeps}a. Note that \cite{dijkstra2015sensitivity} consider a global transition from ice-covered to ocean-covered earth - here we model a large but regional change in ice cover with a smaller contrast in global albedo between the two states; our choice of parameters might be more realistic for albedo variations between glacial and interglacial states. 

We add a stochastic term to (\ref{eq:EBMdet}) that represents unresolved subgrid processes with a fixed amplitude $\eta_T$:
\begin{equation}
\label{eq:EBM}
c_T\, dT= F(T,C) dt + \eta_T dW_T.
\end{equation}
The motivation for noise on the $T$ variable is to model internal variability of climate heat transport processes. The parameters listed in Table~\ref{tab:EBMparams} are used, except where specified. Note that the deterministic equilibria of (\ref{eq:EBMdet}) are at $F(T,C)=0$, which gives
\begin{equation}
C=\Gamma(T):=C_0 \exp\left[\frac{\epsilon(T)\sigma T^4-Q_0(1-\alpha(T))}{A}\right].
\label{eq:EBMequil}
\end{equation}
From (\ref{eq:EBMequil}), this means we have equilibria at
\begin{equation}
\label{eq:EBMDRTeqm}
\Delta R_{[CO_2]}=A\ln(\Gamma(T)/C_0)= \epsilon(T)\sigma T^4-Q_0(1-\alpha(T))
\end{equation}

\begin{figure}[!ht]
\centerline{
\includegraphics[width=10cm]{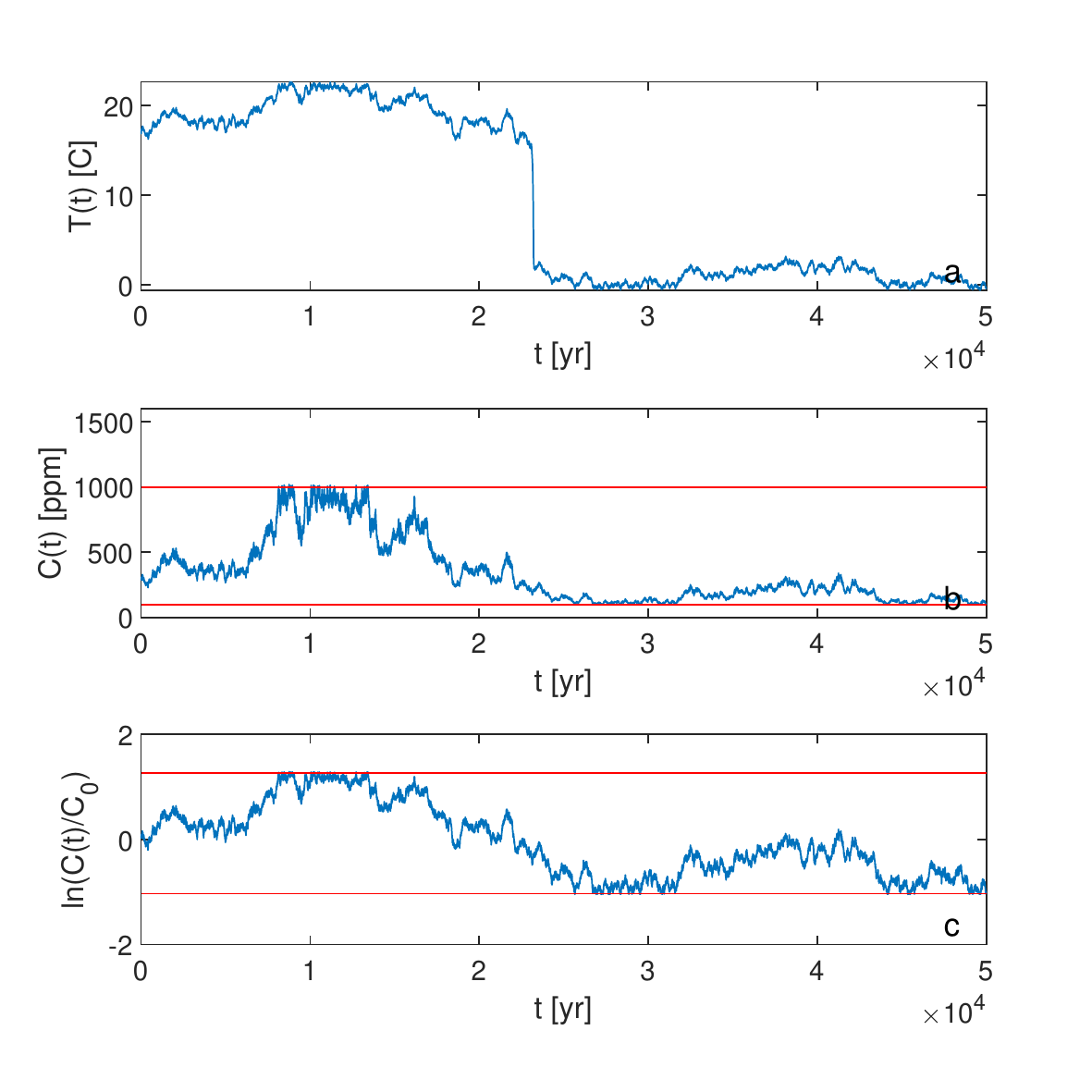}
}
\caption{
Time series for the energy balance model \eqref{e:energybalance} with wandering \coo{} \eqref{eq:coowandering}. (a) Global mean temperature $T(t)$ and (b,c) atmospheric \coo{} $C$ in original (b) and logarithmic coordinates (c). The red lines indicate the limits imposed on the randomly wandering \coo{}. See text for details.}
\label{fig:EBMtimeseries}
\end{figure}

Figure~\ref{fig:alphaeps} illustrates temperature dependence of albedo and emissivity as well as the resulting equilibrium forcing $\Delta R=A \ln (\Gamma(T)/C_0)$ needed to give this temperature. Note there is a unique equilibrium for each $T$, but not necessarily for each $C$: as discussed in \cite{dijkstra2015sensitivity,zaliapin2010another} there are three branches of equilibria for a range of $C$: for the parameters used there is bistability in the region
\begin{equation}
-1.744 ~Wm^{-2}<\Delta R<3.004~Wm^{-2}, \ 202~\ppm{} < C < 490~\ppm{},
\label{eq:bistabrange}
\end{equation}
denoted using the red lines in Figs.~\ref{fig:EBMtimeseries} and \ref{fig:EBMRTearlywarningslowfast}.
We can define the instantaneous sensitivity\footnote{This is referred to as {\em local slope} sensitivity in \cite{Koehler:2017jj}.} as $S=1/\lambda$, where
\begin{equation}
\lambda=\frac{d}{dT} \Delta R_{[CO_2]} = [\epsilon'(T) T+4\epsilon ]\sigma T^3+Q_0\alpha'(T)
\label{eq:iSens}
\end{equation}
is the total feedback factor in this model: $S$ corresponds to the slope of the tangent of the equilibrium (non-stochastic) model (see Fig.~\ref{fig:schem}, point B).  

The sensitivities on the stable branches differ due to both nonlinearity of black body radiation and change in emissivity. Due to the choice of albedo and emissivity changes in this model, $\alpha'$ is nonzero only in the bistable regime and $\epsilon'$ is nonzero only in the temperature range where it varies ($T_0-T_{\epsilon}/2 \leq T \leq T_0+T_{\epsilon}/2$): see Figure~\ref{fig:alphaeps}. 
State-dependence between glacial and interglacial states has been detected in estimates of specific climate sensitivities from different palaeo-data, suggesting lower sensitivity during cold periods than during warm periods (e.g. \cite{Heydt2014,Friedrich:2016dn} who estimate a close approximation of the Charney sensitivity and find warm (interglacial) climate states to be about 60\% more sensitive than cold (glacial) states). 

We can compute the curvature of (\ref{eq:EBMDRTeqm}) as
$$
\frac{d^2}{dT^2} \Delta R_{[CO_2]} = (\epsilon''T^2+ 8\epsilon T+12\epsilon')T^2-Q_0\alpha''
$$
Note that this is small except near the folds at $T\approx T_1$ and $T\approx T_2$ with maximum absolute value $
\frac{d^2}{dT^2} \Delta R$ of order $T_{\alpha}^{-1}$. This confirms that the saddle-node becomes non-smooth in the limit $T_{\alpha}\rightarrow 0$. The second derivative gives the size of the quadratic correction $a$ in Zaliapin et al. \cite{zaliapin2010another}; very large values of the slope near the saddle nodes correspond to the run-away climate observed in \cite{BlochJohnson:2015hm}.

For the model (\ref{eq:EBM}), the atmospheric \coo{}  concentration is a parameter for the energy balance dynamics. We explore this by considering a `wandering' \coo{} profile such that $\gamma(t):=\ln(C(t))$ undertakes a Brownian motion with growth in variance $\eta_{\gamma}$ per unit time between reflecting limit values. More precisely, we consider \coo{} dynamics governed by soft reflecting boundary conditions at $\ln C_{min}$ and $\ln C_{max}$:
\begin{equation}
d\gamma= K \theta(\gamma)\,dt+ \eta_{C}\,dW_\gamma
\label{eq:coowandering}
\end{equation}
where the noise in the $C$ variable (with fixed amplitude $\eta_C$) represents variability in $\coo$ forcing. We assume
\begin{equation}
\theta(\gamma):=H(\ln C_{min}-\gamma)(\ln C_{min}-\gamma)+H(\gamma-\ln C_{max})(\ln C_{max}-\gamma)
\label{eq:FC}
\end{equation}
and we use parameters
\begin{equation}
K=10^{-7} ~s^{-1},~C_{min}=10^2 ~\ppm,~~C_{max}=~10^3 ~\ppm, ~ \eta_{C}=2\times 10^{-6} ~s^{-1/2}.
\end{equation}
Clearly there are common causes of variability of temperature and \coo{} and so in general there will be strong correlations between the noise terms $W_T$ and $W_\gamma$; for convenience we assume here that they are uncorrelated. 
In most studies of climate sensitivity, carbon cycle processes are not treated as feedbacks but rather as forcing (external to the system); this assumption corresponds in our model to \coo{} driving temperature changes with no direct impact of temperature on \coo{}. The parameter $\eta_{C}$ determines the timescale of wandering of the \coo{}: we consider cases where this is slower than, or comparable to the the timescale of evolution of $T$.

\begin{figure}[!ht]
\centering
\includegraphics[width=12cm]{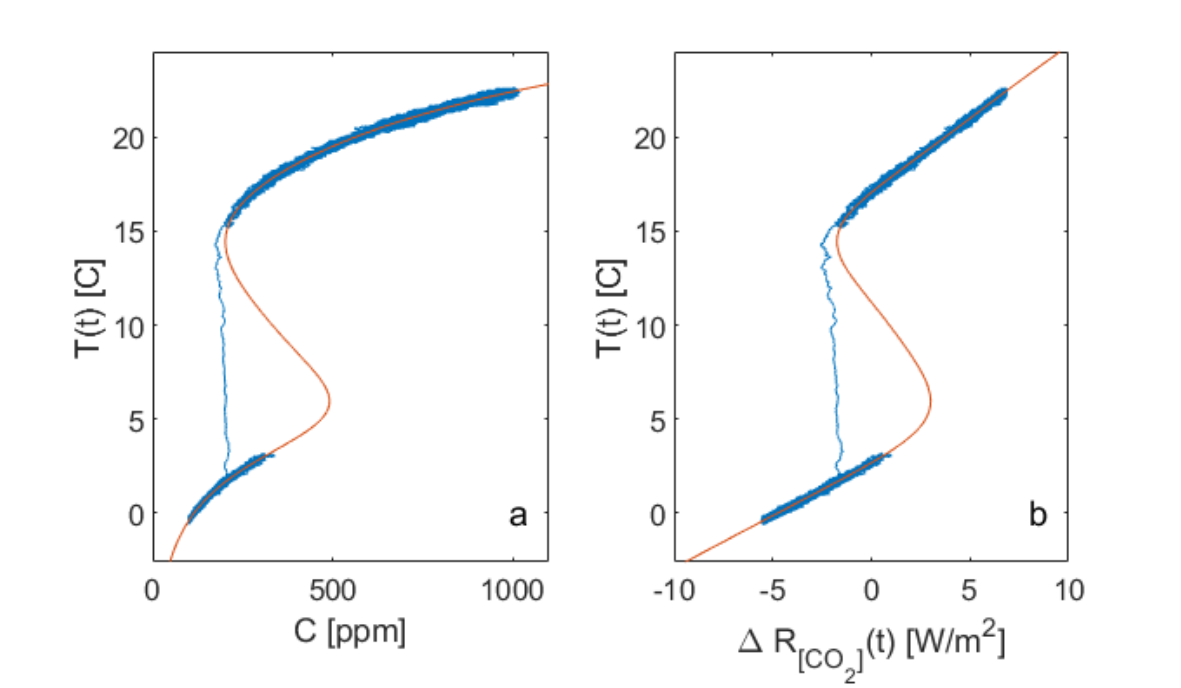}

\caption{(a) Temperature versus (a) atmospheric \coo{} concentration and (b) $\Delta R$ of radiative forcing by \coo{}, for the time series in Figure~\ref{fig:EBMtimeseries}. The red line corresponds to the steady solution of (\ref{eq:EBMdet}) with dependence on \coo{}.}
\label{fig:EBMTRplot}
\end{figure}

Figure~\ref{fig:EBMtimeseries}(b) shows a time series for a typical simulation of (\ref{eq:EBM}) with wandering \coo{} (\ref{eq:coowandering}) and parameters as in Table~\ref{tab:EBMparams}, while the corresponding time series of $T$ is shown in Figure~\ref{fig:EBMtimeseries}(a);
we see as $C$ crosses thresholds (for the bistable regime as calculated in (\ref{eq:bistabrange})) the state of the system tips between warm and cold states. Global mean temperature $T$ vs $C$ and the relative radiative forcing $\Delta R_{[CO_2]}$ are shown in Figure~\ref{fig:alphaeps} (c,d) for the non-stochastic model and in Figure~\ref{fig:EBMTRplot} for the stochastic model, respectively. Observe the region of bistability around $\Delta R_{[CO_2]}=0$ (Figure~\ref{fig:EBMTRplot}b) that switches rapidly between cool high albedo and warm low albedo states via saddle-node bifurcations. There is approximate linearity away from these tipping points, but with different mean slopes.

\begin{table}[ht!]
    \centering
    $$
    \begin{array}{|c|rl||c|rl|}
    \hline
        \alpha_1 & 0.52 & - & \alpha_2 & 0.47 & - \\
        T_1 & 278 & K & T_2 & 288 & K \\
        \epsilon_1 & 0.53 & - & \epsilon_2 & 0.39 & - \\
        A & 5.35 & W m^{-2}& T_0 & 288 & K \\
        T_{\alpha} & 5 & K & T_{\epsilon} & 20 & K\\
        C_0 & 280 & \ppm{} & Q_0 & 342 & W m^{-2}\\
        C_T & 5 \times 10^{8} & J m^{-2} K^{-1} & \sigma & 5.67 \times 10^{-8} & W m^{-2} K^{-4}\\
        \eta_T & 5\times 10^{-6} & K s^{-1/2} & \eta_C & 2\times 10^{-6} & s^{-1/2}\\
    \hline
    \end{array}
    $$
    \caption{Parameters for the energy balance model (\ref{eq:EBMdet},\ref{eq:EBM}) adapted from \cite{dijkstra2015sensitivity} to include state-dependent emissivity. Note that \cite{dijkstra2015sensitivity} consider a global transition and so use different values: $\alpha_1=0.7$ $\alpha_2=0.2$, $T_1=263~K$, $T_2=293~K$, $T_{\alpha}=0.273 K$, $A=20.5~Wm^{-2}$ and have an additional constant $150~Wm^{-2}$.}
    \label{tab:EBMparams}
\end{table}

\subsection{Extreme sensitivities, early warning signals and tipping between regimes}

The energy balance model (\ref{eq:EBM}) with wandering \coo{} (\ref{eq:coowandering}) gives a framework in which one can test correlation between extreme values of sensitivity and tipping between climate regimes, as well as testing other possible early warning signals for a tipping event. Indeed, we find a rise in instantaneous sensitivity seems to act as a good precursor in cases of slow variation of \coo{}.

\begin{figure}[!ht]
\centering
\includegraphics[width=5.5cm]{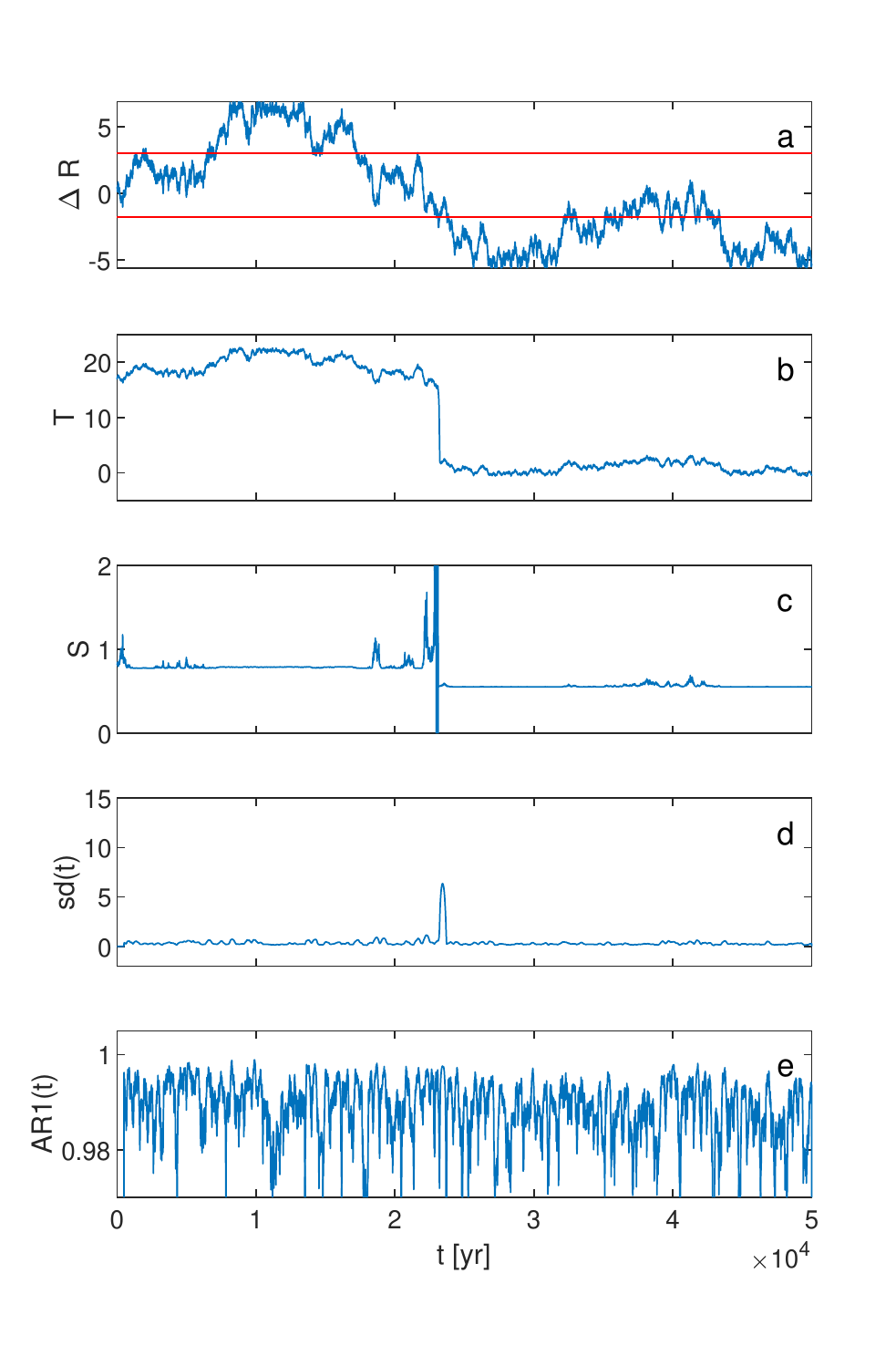}
\includegraphics[width=5.5cm]{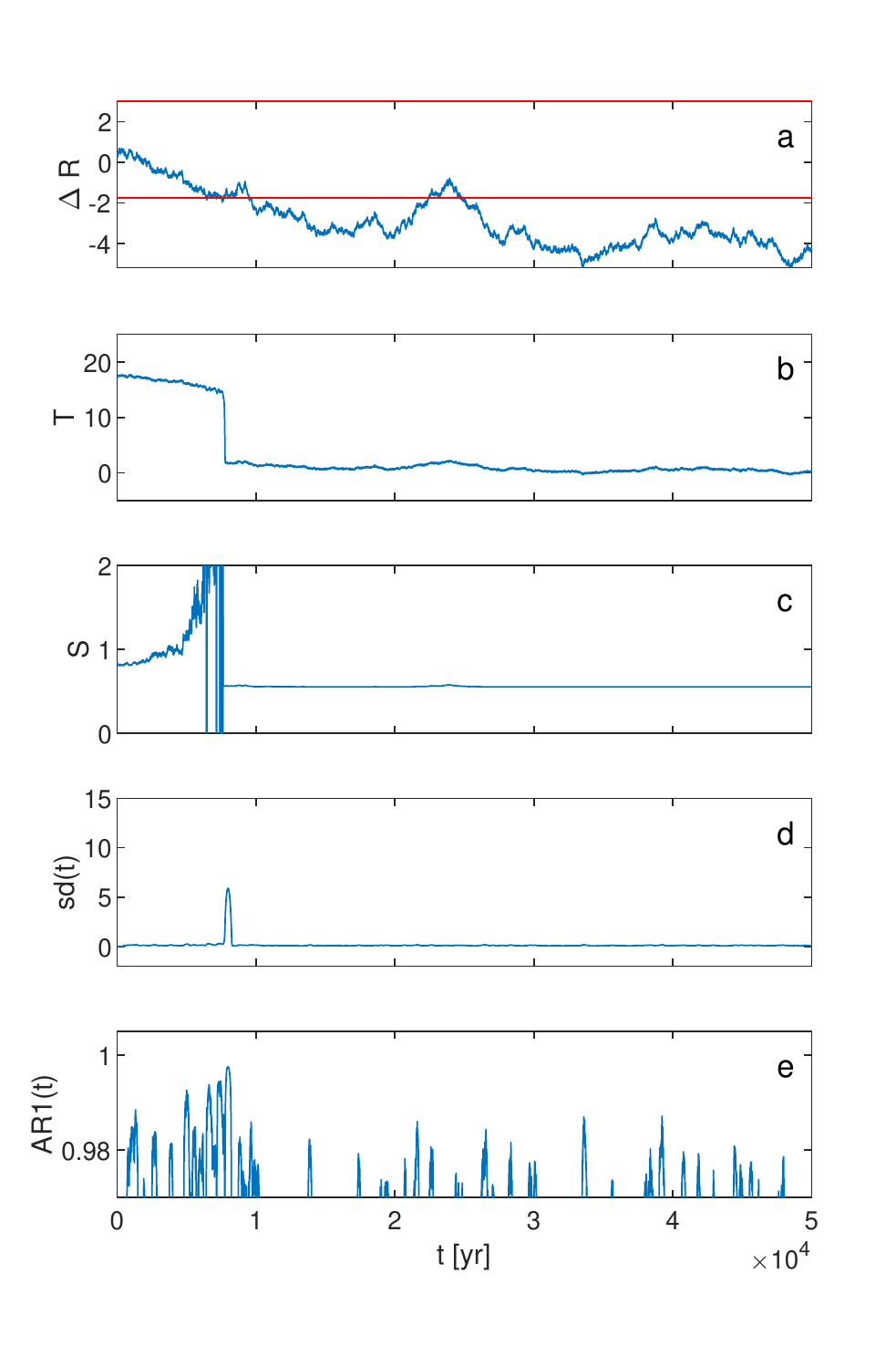}

\caption{Examples of tipping events for $T$ for the energy balance model (\ref{eq:EBMdet},\ref{eq:EBM}) with wandering \coo{} (\ref{eq:coowandering}) with (left) parameters as in Figure~\ref{fig:EBMtimeseries}, i.e. with $(\eta_T,\eta_C)=(5\times 10^{-6},2\times 10^{-6})$ and (right) slower variation of $C$ $(\eta_T,\eta_C)=(5\times 10^{-6},5\times 10^{-7})$. (a,b) time series of $\Delta R$ and $T$. The red lines on (a) show the locations of the saddle-node bifurcations that bound the region of bistability (\ref{eq:bistabrange}). (c) shows estimated instantaneous sensitivity from the nearest equilibrium of (\ref{eq:EBM}). Note the gradual rise and fluctuations in $S$ on approach to the tipping point, and the two levels of $S$ corresponding to the differing sensitivities of the two stable branches. (d,e) show standard deviation and AR1 coefficient: note that the AR1 coefficient seems to have predictive power only for the right column.
}
\label{fig:EBMRTearlywarningslowfast}
\end{figure}

Figure~\ref{fig:EBMRTearlywarningslowfast}(left) shows the variation of instantaneous sensitivity and two early warning signals for tipping between regimes for the wandering variation of \coo{} concentration $\ln C(t)$ relative to the timescale of evolution of $T$, with $\eta_C=2\times 10^{-6}~ s^{-1/2}$; this means that \coo{} variations are comparable in timescale to $T$ fluctuations. By considering the nearest equilibrium point on $C=\Gamma(T)$ for fixed $C$ (\ref{eq:EBMequil}), we evaluate the instantaneous sensitivity $S$ using (\ref{eq:iSens}) and plot this in middle panel. 
Observe there is a qualitative change in $S$ before and after tipping events. There is a clear precursor and then singularity as the tipping point is crossed. Note that in this case the instantaneous sensitivity depends only on the current \coo{} level and the nearest branch - fluctuations in $T$ around the branch do not affect $S$, while fluctuations in $C$ do. There are also apparent `false alarms': for example, the fluctuations of $S$ around 3~kyr and 18~kyr. 
Figure~\ref{fig:EBMRTearlywarningslowfast}(left, panels d,e) show detrended estimates of $sd(t)$ and $AR1(t)$ \cite{Zhang:2015} using moving averages with length $\tau=500$yrs. Neither show any clear precursors before tipping events. By contrast, Figure~\ref{fig:EBMRTearlywarningslowfast}(right) shows early warning signals for tipping between regimes for slower variation of \coo{} concentration with $\eta_C=5\times 10^{-7}~ s^{-1/2}$, where $T$ evolves faster than $C$. Unlike the case on the left, this slower switching gives a visible precursor of increasing $AR1$. In both cases there are increasing fluctuations of instantaneous sensitivity $S$. Note however, that the instantaneous $S$ we consider here uses the model equations, and hence will be more difficult to access from complex model realisations or observations.

\subsection{Two-point sensitivities and tipping}

\begin{figure}[!ht]
\centering
\includegraphics[width=8cm]{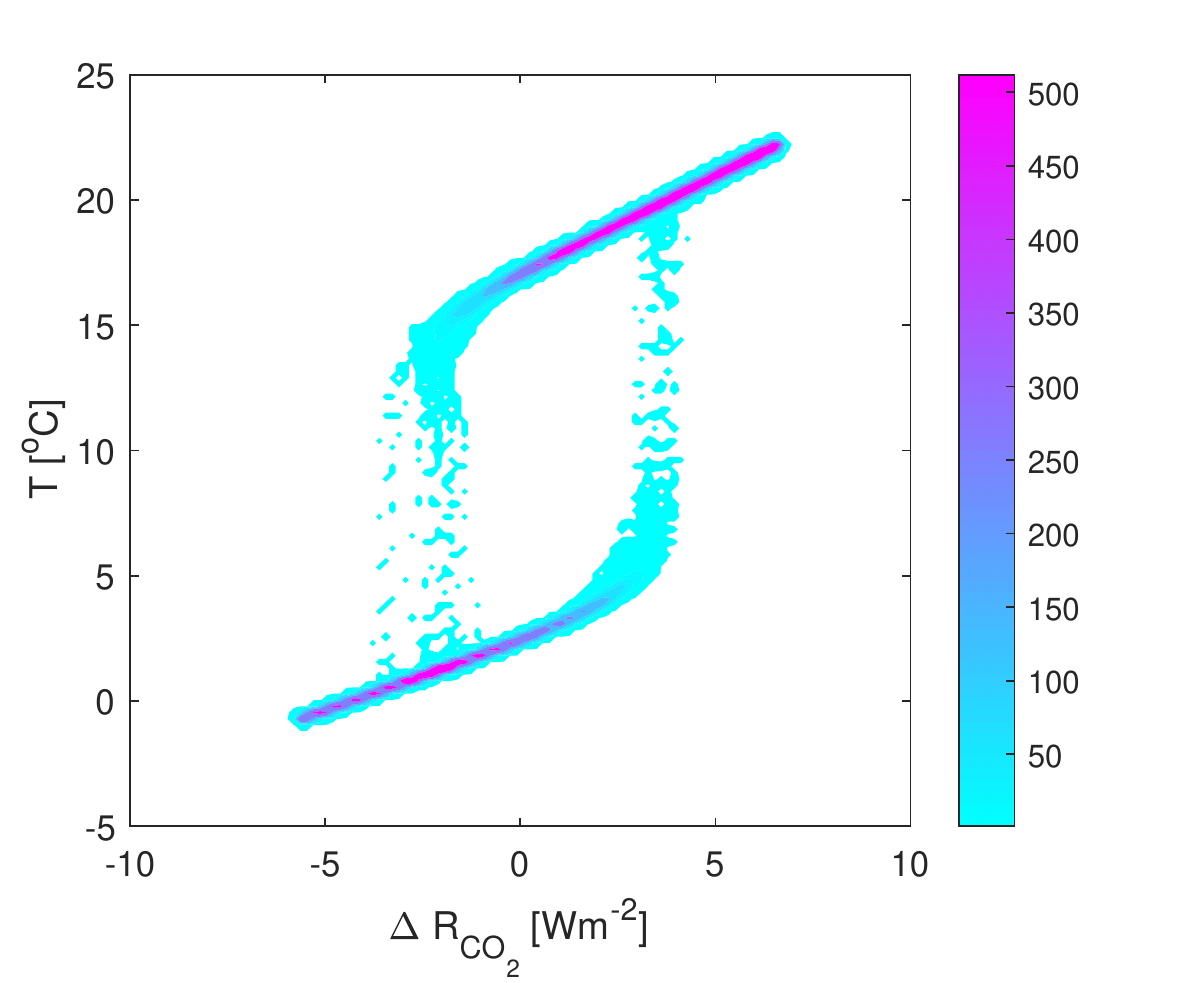}

\caption{Density of $T$ versus $\Delta R=A\ln (C/C_0)$ of radiative forcing by \coo{} for a longer time series similar to that in Figure~\ref{fig:EBMTRplot}(b). Observe the two regions with approximately linear relationship (cf Figure~\ref{fig:schem}): the colour scale indicates density [arbitrary units].}
\label{fig:EBMRTplot}
\end{figure}

\begin{figure}[!ht]
\centering
\includegraphics[width=12cm]{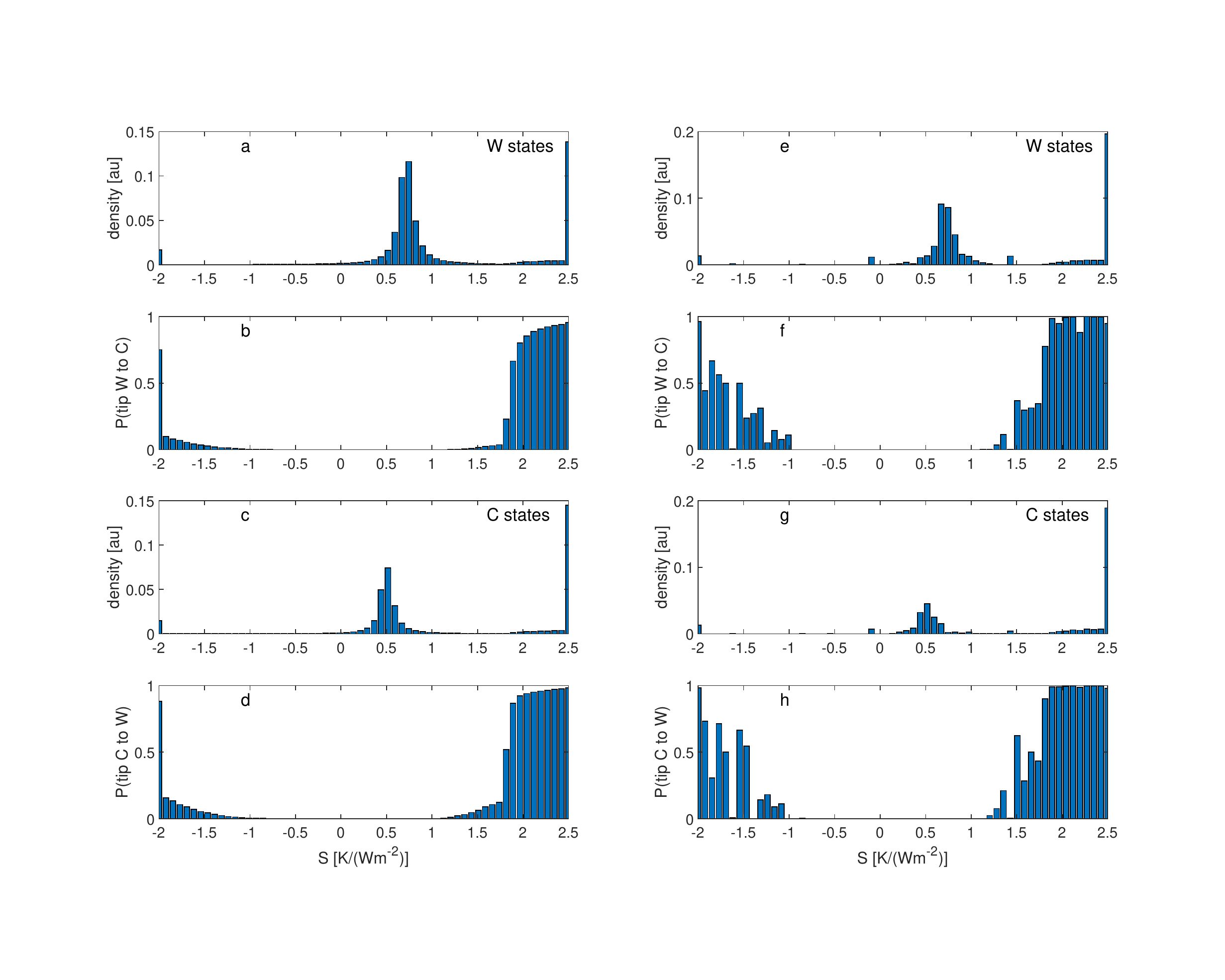}

\caption{(a,c,e,g) Conditional two-point sensitivities and (b,d,f,h) probabilities of tipping from warm (a,b,e,f) and cold (c,d,g,h) states, for the energy balance model (\ref{eq:EBM}) with wandering \coo{} (\ref{eq:coowandering}). (a-d) are computed using a range of delays up to $20~kyr$ while (e-h) are computed using $10^7$ independently sampled pairs of points from the distribution in Figure~\ref{fig:EBMRTplot}. Note that sensitivities outside the horizontal range are rounded to the last bin.
}
\label{fig:EBMRTsensitivities}
\end{figure}

An approximation of the stationary density of the global attractor for the system (\ref{eq:EBM},\ref{eq:coowandering}) is shown in Figure~\ref{fig:EBMRTplot}. We classify the system regime as one of:
\begin{equation*}
\left\{
\begin{array}{lr}
\mbox{Warm (W)} & \mbox{ if }T>T_{\mathrm{thr}}\\
\mbox{Cold (C)} & \mbox{ if }T\leq T_{\mathrm{thr}}
\end{array}\right.
\end{equation*}
where we choose a threshold $T_{\mathrm{thr}}=10~C$ between the two stable branches: see Figure~\ref{fig:schem}. We simulate a single very long trajectory ($5\times 10^5$ years) of the energy balance model (\ref{eq:EBMdet},\ref{eq:EBM}) with wandering \coo{} and use this to create a density plot of the climate attractor projected onto $T$ vs $\Delta R$, as shown in Figure~\ref{fig:EBMRTplot}. This is used to consider the two-point sensitivities and probabilities of tipping as in Figure~\ref{fig:EBMRTsensitivities}. Panels (a-d) are computed by sampling incremental sensitivities (\ref{eq:Sinc}) from points for  increments up to $20~kyr$. Panels (e-h) are computed by sampling $10^7$ pairs of points from the distribution in Figure~\ref{fig:EBMRTplot} using the two-point sensitivity (\ref{eq:S01}).  We observe:
\begin{itemize}
    \item There is good qualitative agreement between the incremental sensitivities averaged over long delays and the two-point sensitivities sampled independently from the attractor. Indeed, the autocorrelation of the timeseries for $T$ (not shown) has substantially decayed and has its first zeros around $20 kyr$.
    \item High probability of tipping (see Figure~\ref{fig:EBMRTsensitivities}(b,d)) corresponds mostly to extremes of $S$ that may be positive or negative $S$.
    \item Within the W and C regimes, the sensitivities are closely clustered but have different means for the W and C state. We can estimate these using average temperatures and (\ref{eq:iSens}) as $S\approx 0.79 ~K[Wm^2]^{-1})$ for the W and $S\approx 0.55$ for the C state, respectively. 
    \item Note that there are relatively low-probability `shoulders' of the distributions within-regime. These are due to the classification of regimes also including states that are in transitions: although the system is in transition, both points are still classified as the same regime.
    \item A hysteresis-like effect may appear even in the absence of bistability. If the dynamic change in a system parameter is fast enough then an apparent hysteresis may arise due to different lags for a rising or falling parameter: see for example \cite{herein2016}. However, Figure~\ref{fig:EBMTRplot} suggests this is not the case here.
\end{itemize}

\subsection{Sensitivities in the absence of tipping}

\begin{figure}
    \centering
    \includegraphics[width=5.5cm]{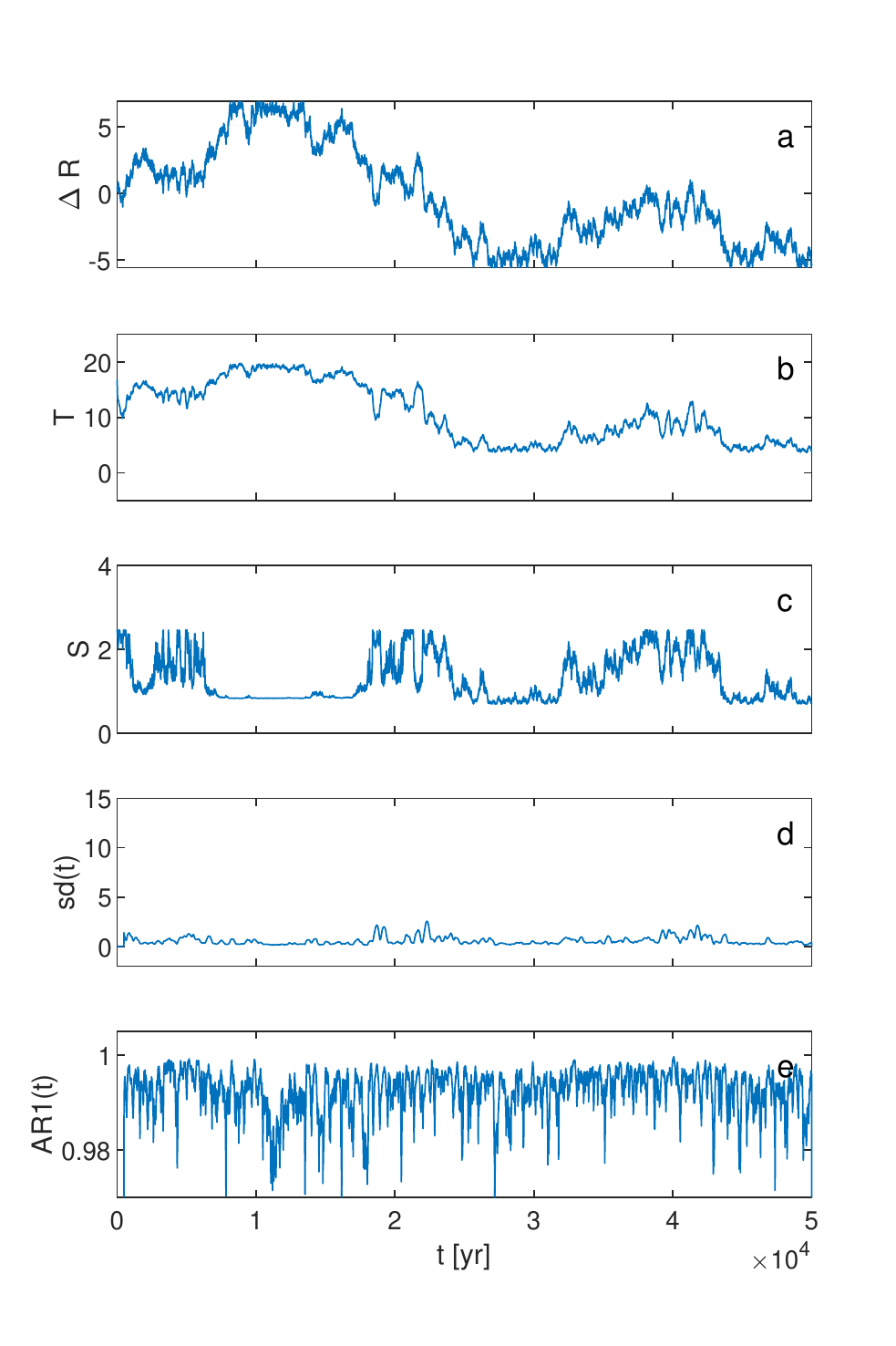}
    \includegraphics[width=5.5cm]{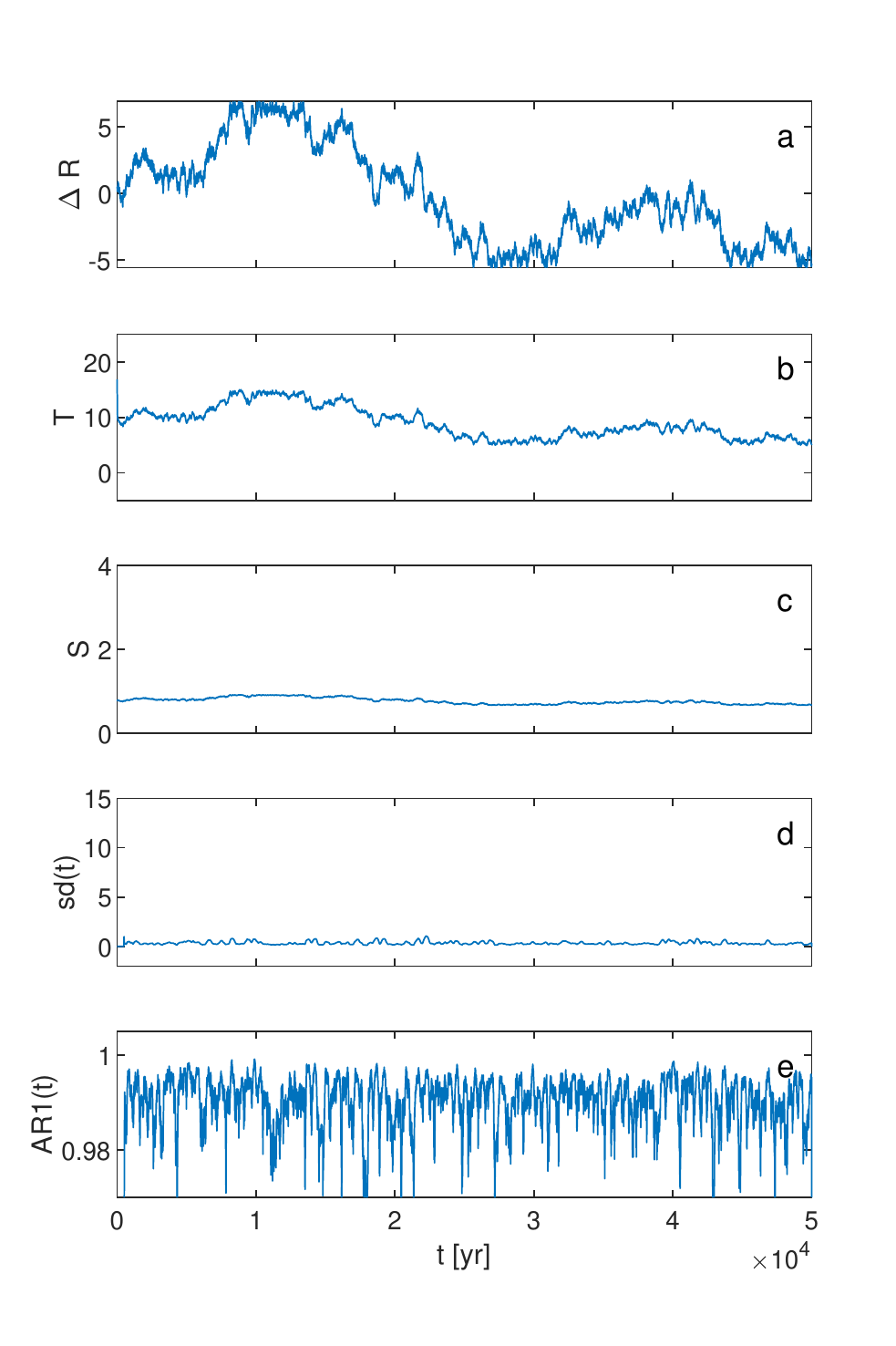}
    
    \caption{Response of global mean temperature to the same realisation of wandering \coo{} for the energy balance model (\ref{eq:EBM}) with (left) low albedo contrast and (right) no albedo contrast; (a-e) as in Figure~\ref{fig:EBMRTearlywarningslowfast}. In both cases there is no region of bistability. Observe there are large but bounded fluctuations of the instantaneous sensitivity in case (left), indicating state-dependency but no tipping. In case (right) there is comparatively little fluctuation of sensitivity; compare with Figure~\ref{fig:EBMRTearlywarningslowfast}(left) for the default parameters with a region of bistability.}
    \label{fig:albedochanges}
\end{figure}

\begin{figure}
    \centering
    \includegraphics[width=11cm]{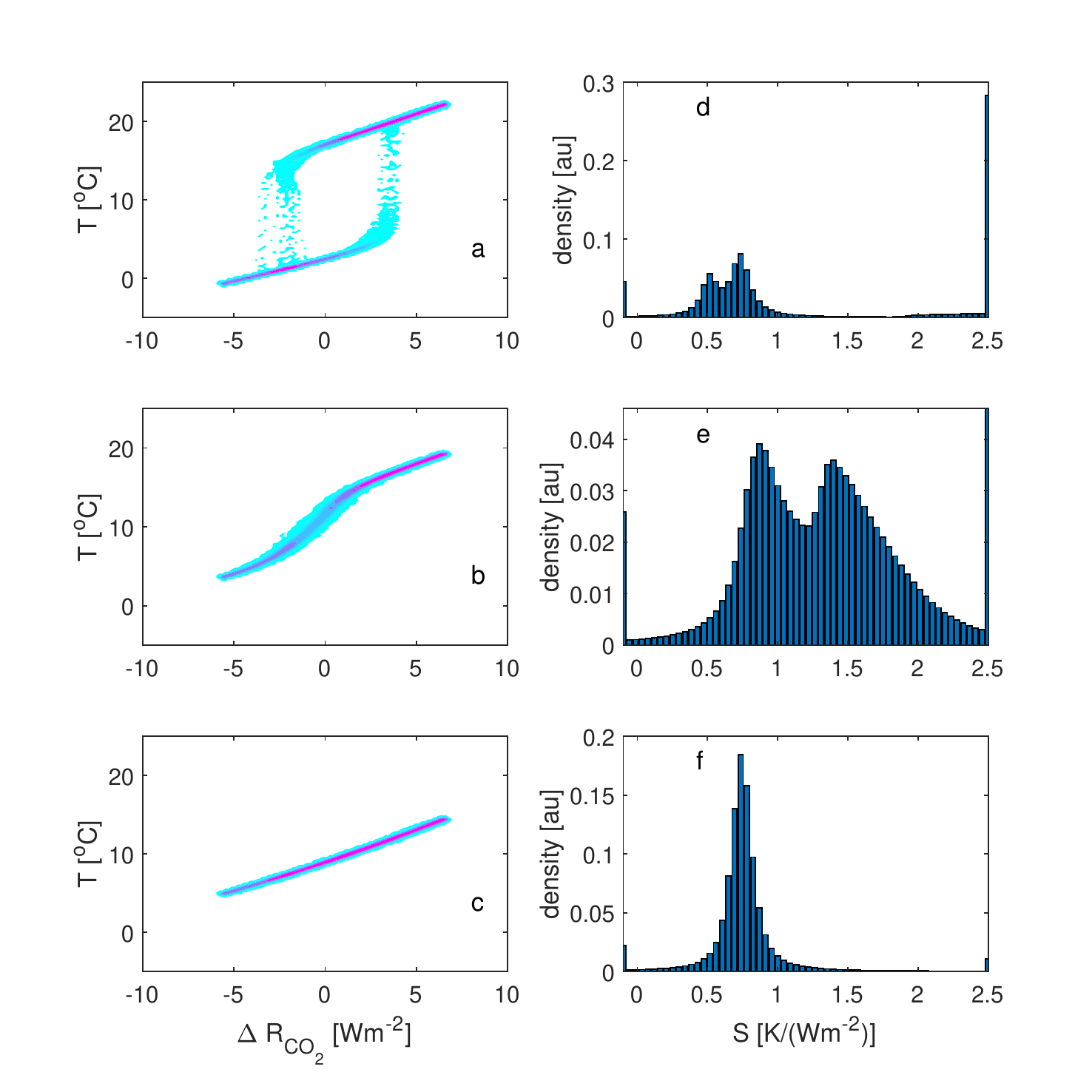}
    \caption{(a-c) Plots of $T$ vs $\Delta R$ for a long timeseries of the energy balance model (\ref{eq:EBM}) with wandering \coo{}, parameters as in Figure~\ref{fig:EBMRTplot}. (d-f) plots of distributions of two-point sensitivities corresponding to (a-c) respectively. (a,d) correspond to the default scenario with bistability, (b,e) to the low albedo contrast and (c,f) to the no albedo contrast scenario; cf Figure~\ref{fig:albedochanges}. Observe the presence of two peaks (corresponding to state dependent sensitivities on the two branches) and a large mass of extremes for (d). (e) shows a range of state-dependent variation but relatively few extremes; while (f) resembles a normal distribution. Observe that (d) can be decomposed into the conditional sensitivities shown in Figure~\ref{fig:EBMRTsensitivities}.}
    \label{fig:albedochangesRT}
\end{figure}

Considering the same model (\ref{eq:EBMdet},\ref{eq:EBM}) with wandering \coo{} (\ref{eq:coowandering}), we use different parameters to contrast the results in the previous sections with cases where there is no bistability. In particular we consider parameters as in Table~\ref{tab:EBMparams} except for:
\begin{itemize}
    \item Default albedo contrast: $\alpha_1=0.52$, $\alpha_2=0.47$ (i.e. also as in Table~\ref{tab:EBMparams}). 
    \item Low albedo contrast: $\alpha_1=0.50$, $\alpha_2=0.48$.
    \item No albedo contrast: $\alpha_1=\alpha_2=0.495$.
\end{itemize}
Figure~\ref{fig:albedochanges} shows the low and no albedo contrast cases, comparing to the default case Figure~\ref{fig:EBMRTearlywarningslowfast}(left). For the low albedo contrast case there is no longer a region of bistability, but there is nontrivial variation of the instantaneous sensitivity along the attractor. For the no albedo contrast case, the instantaneous sensitivity is close to constant. The projection of the climate attractor into the $(\Delta R,T)$ plane is shown in Figure~\ref{fig:albedochangesRT}(a-c), while the corresponding distribution of two-point sensitivities in Figure~\ref{fig:albedochangesRT}(d-f). Observe the presence of non-unimodal distributions for (d,e) associated with regions with different two-point  sensitivity, and clear skewness and tails again associated with the geometry of the measure in (a,b). Note the higher average in (e) corresponds to there being only a single regime in (b) that runs over a wide range of temperatures.

In physical terms, the skewness (and long tails) in (d,e) originate from the state-dependence and nonlinearity of feedbacks (i.e. non-constant feedback factors). The bistability of the two regimes with different feedbacks gives the two peaks in the distribution of Figure~\ref{fig:albedochangesRT}(d). However, Figure~\ref{fig:albedochangesRT}(e) still has two peaks: these originate from state dependence on the same attractor (Figure~\ref{fig:albedochangesRT}(b). For this low-albedo-contrast case, there is no `tipping' but we still find very non-Gaussian distribution of $S$ that comes from nonlinearities in the system that, in this case, do not produce tipping. Note that only in the no albedo contrast case (c) is there a plausible fit to Gaussian.

\section{Discussion}
\label{sec:discuss}

We demonstrate that state-dependence and the presence of tipping points produces signatures in the distribution of instantaneous and two-point notions of ECS. We explore this using a global energy balance model where state-dependence and multistability originate from the dependence of both albedo and emissivity on temperature.

For the deterministic version of our model (\ref{eq:EBMdet}) with fixed \coo{} the changes in albedo mean there can be bistability between regimes, while the changes in emissivity contribute to different sensitivities within these regimes. The distribution of ECS comes from several sources - nonlinearities that result in tipping points and/or state-dependence of the feedbacks and sub-grid variability that we model here as stochastic perturbations. Such regime-dependent sensitivity and extremes associated with tipping points are also visible in the more complex Gildor and Tziperman model \cite{Gildor:2001,Heydt:2017}, as outlined in Appendix~\ref{sec:GT}.

For the stochastic model (\ref{eq:EBM}) with wandering \coo{}, regime-dependent sensitivity is visible as differences in slope of the stable regimes for the $T$ vs $\Delta R$ plots (see Figure~\ref{fig:EBMTRplot}(b)). 
The densities of the stable regimes for the $T$ vs $\Delta R$ plots (see Figure~\ref{fig:EBMRTplot}) show varying slopes and so conditional two-point sensitivities for the two regimes (see e.g. Figure~\ref{fig:EBMRTsensitivities}(a) and (c)) can have peaks at different sensitivities. 
We compare several notions of sensitivity. These are the instantaneous sensitivity associated with the slope of the equilibrium branch, incremental sensitivity associated with a fixed delay, and two-point sensitivity that compares arbitrary points on the climate attractor. The presence of tipping points gives extremes of sensitivities in that (i) there are large fluctuations of instantaneous sensitivities for the nearest equilibrium just before a tipping point (see Figure~\ref{fig:EBMRTearlywarningslowfast}); and (ii) in the distributions of conditional two-point sensitivities that cross regimes (see Figures~\ref{fig:EBMRTsensitivities}(d) and (f)). It is remarkable that the two-point sensitivities are so informative, given that they compare points that may be very far from each other in time. 

There remains much work to be done to understand the relation between and limitations of these notions of ECS, and indeed ECS calculated in other ways, for example from instantaneous doubling of \coo{}: this will involve transient non-equilibrium processes due to ocean thermal inertia. Note that determining ECS from palaeoclimate time-series \cite{Heydt2016rev}, the two-point notion clearly has an advantage that we are not limited by the time-resolution of the time series.

The energy balance model can be criticised as being very simple and hard to parametrize in terms of the various physical processes that contribute to albedo and emissivity. Moreover, we consider \coo{} in (\ref{eq:EBM}) purely as a forcing term which ignores known land surface and ocean processes where temperature is known to affect \coo{} balance.  However, the model is complex enough to confirm that extremes in ensembles of computed climate sensitivities can indicate nearby tipping points. Computations presented in Appendix~\ref{sec:GT} confirm this picture in a more complex box-model for the glacial cycles, where the \coo{} is modelled dynamically. 

\subsection{Future perspectives}

When (and how) extremes of sensitivity can be effective precursors of a tipping event will depend on a number of factors. In particular, the timescale of dynamics of the climate response needs to be faster than the timescale of changes in forcing.   Figure~\ref{fig:EBMRTearlywarningslowfast}(left) shows that as \coo{} variability is rapid, this results in tipping points with little precursor visible in changes to $AR1$, though it is visible in the instantaneous sensitivity. 
There may be `rate-induced' tipping points \cite{Ashwinetal:2017,Ashwinetal:2012} that appear when the timescale of the forcing interacts with that of the system. The size of the region of effective nonlinearity can also vary. For (\ref{eq:EBM}) this is affected by the temperature scales $T_{\alpha}$ and $T_{\epsilon}$ over which the albedo and emissivity changes occur.  Note also that although tipping points do give rise to extremes in the distribution of ECS, extremes do not necessarily indicate a tipping point.

Translating these results to a time-dependent setting and to more complex models will be difficult: the possible states in the `climate attractor' and the associated invariant measure $\mu$ is harder to define in the presence of non-stationary temporal variation of forcing, or for large recurrence times and a variety of nonlinear multiscale processes. In such cases, interpreting transitions as tipping points is a challenge; nonetheless, the palaeoclimate record does show a variety of large and sudden transitions \cite{Lenton:2011}. For example, ice core/ocean core records indicate repeated sudden changes in (regional) surface temperature associated with glacial cycles \cite{hogg2008glacial} or Dansgaard-Oeschger \cite{ditlevsen2005recurrence} events as well as global transitions, for example the greenhouse-icehouse transition at the Eocene-Oligocene transition \cite{Coxall:2005}. Although glacial cycles can be found in models such as \cite{Gildor:2001} as relaxation oscillation with clear regimes, for climate reconstruction data these regimes are not so clear (e.g. \cite{Koehler:2017jj,Friedrich:2016dn}). 

Recent work \cite{Steffen:2018} suggests we are at a crossroads in terms of the future earth system state. On the one hand, looking at the palaeoclimate record for the last 1 million years suggests that we are overdue descent into an ice age. On the other hand, comparison of anthropogenic \coo{} emissions with the palaeo record suggest the next tipping point may be to much warmer `hothouse' earth. A better understanding of improved indicators such as two-point ECS and what they say about the climate response to changes in greenhouse gases, together with a better understanding of hothouse earth climate states that may have existed in the past (e.g. the Palaeocene climate \cite{Huber2011,Baatsen:2018hz}) should help our understanding and guide future generations in their need to avoid dangerous climate change.

\subsection*{Acknowledgements}
We thank Peter Cox, Henk Dijsktra, Valerio Lucarini, Michael Ghil and Georg Gottwald for interesting comments and suggestions concerning this work. We thank the EU ITN network CRITICS and CliMathNet for opportunities to discuss and develop this research. AvdH thanks the program of the Netherlands Earth System Science Centre (NESSC), financially supported by the Dutch Ministry of Education, Culture and Science (OCW). This project is TiPES contribution \#7: This project has received funding from the European Union's Horizon 2020 research and innovation programme under grant agreement No 820970.

%
\subsection*{Conflict of interest}
The authors declare that they have no conflict of interest.

\bibliographystyle{plain}

\begin{thebibliography}{10}
	
	\bibitem{Ashwinetal:2017}
	Peter Ashwin, Clare Perryman, and Sebastian Wieczorek.
	\newblock Parameter shifts for nonautonomous systems in low dimension:
	bifurcation-and rate-induced tipping.
	\newblock {\em Nonlinearity}, 30(6):2185, 2017.
	
	\bibitem{Ashwinetal:2012}
	Peter Ashwin, Sebastian Wieczorek, Renato Vitolo, and Peter Cox.
	\newblock Tipping points in open systems: bifurcation, noise-induced and
	rate-dependent examples in the climate system.
	\newblock {\em Philosophical Transactions of the Royal Society A: Mathematical,
		Physical and Engineering Sciences}, 370(1962):1166--1184, 2012.
	
	\bibitem{Baatsen:2018hz}
	Michiel L~J Baatsen, Anna~S von~der Heydt, M~Huber, Michael~A Kliphuis, Peter~K
	Bijl, Appy Sluijs, and Henk~A. Dijkstra.
	\newblock {Equilibrium state and sensitivity of the simulated middle-to-late
		Eocene climate}.
	\newblock {\em Clim. Past Disuss.}, in review, 2018.
	
	\bibitem{BlochJohnson:2015hm}
	Jonah Bloch-Johnson, Raymond~T. Pierrehumbert, and Dorian~S. Abbot.
	\newblock {Feedback temperature dependence determines the risk of high
		warming}.
	\newblock {\em Geophysical Research Letters}, 42:4973--4980, 2015.
	
	\bibitem{budyko1969effect}
	Mikhail~I. Budyko.
	\newblock The effect of solar radiation variations on the climate of the earth.
	\newblock {\em tellus}, 21(5):611--619, 1969.
	
	\bibitem{Caballero2013}
	Rodrigo Caballero and M~Huber.
	\newblock {State-dependent climate sensitivity in past warm climates and its
		implications for future climate projections}.
	\newblock {\em Proceedings of the National Academy of Sciences},
	110(35):14162--14167, 2013.
	
	\bibitem{Charney:1979}
	J.~Charney.
	\newblock {\em Carbon Dioxide and Climate: A Scientific Assessment}.
	\newblock The National Academies Press, Washington, DC, 1979.
	
	\bibitem{Chekroun:2011}
	Micka{\"{e}}l~D. Chekroun, Eric Simonnet, and Michael Ghil.
	\newblock Stochastic climate dynamics: Random attractors and time-dependent
	invariant measures.
	\newblock {\em Physica D: Nonlinear Phenomena}, 240(21):1685 -- 1700, 2011.
	
	\bibitem{Cox:2018}
	Peter~M. Cox, Chris Huntingford, and Mark~S. Williamson.
	\newblock Emergent constraint on equilibrium climate sensitivity from global
	temperature variability.
	\newblock {\em Nature}, 553(7688):319, 2018.
	
	\bibitem{Coxall:2005}
	H.K Coxall, P.A. Wilson, H.~P\"{a}like, C.H. Lear, and J.~Backman.
	\newblock Rapid stepwise onset of antarctic glaciation and deeper calcite
	compensation in the pacific ocean.
	\newblock {\em Nature}, 433:53--57, 2005.
	
	\bibitem{Dekker:2018bs}
	Mark~M. Dekker, Anna~S. von~der Heydt, and Henk~A. Dijkstra.
	\newblock {Cascading transitions in the climate system}.
	\newblock {\em Earth System Dynamics}, 9(4):1243--1260, 2018.
	
	\bibitem{dijkstra2015sensitivity}
	H.A. Dijkstra and J.P. Viebahn.
	\newblock Sensitivity and resilience of the climate system: A conditional
	nonlinear optimization approach.
	\newblock {\em Communications in Nonlinear Science and Numerical Simulation},
	22(1-3):13--22, 2015.
	
	\bibitem{ditlevsen2005recurrence}
	Peter~D. Ditlevsen, Mikkel~S. Kristensen, and Katrine~K. Andersen.
	\newblock The recurrence time of dansgaard--oeschger events and limits on the
	possible periodic component.
	\newblock {\em Journal of Climate}, 18(14):2594--2603, 2005.
	
	\bibitem{Drijfhout:2015hj}
	Sybren Drijfhout, Sebastian Bathiany, Claudie Beaulieu, Victor Brovkin, Martin
	Claussen, Chris Huntingford, Marten Scheffer, Giovanni Sgubin, and Didier
	Swingedouw.
	\newblock {Catalogue of abrupt shifts in Intergovernmental Panel on Climate
		Change climate models.}
	\newblock {\em Proceedings of the National Academy of Sciences of the United
		States of America}, 112(43):E5777--86, 2015.
	
	\bibitem{Friedrich:2016dn}
	Tobias Friedrich, Axel Timmermann, Michelle Tigchelaar, Oliver~Elison Timm, and
	Andrey Ganopolski.
	\newblock {Nonlinear climate sensitivity and its implications for future
		greenhouse warming}.
	\newblock {\em Science Advances}, 2(11):e1501923--e1501923, 2016.
	
	\bibitem{ghil1976climate}
	Michael Ghil.
	\newblock Climate stability for a sellers-type model.
	\newblock {\em Journal of the Atmospheric Sciences}, 33(1):3--20, 1976.
	
	\bibitem{Gildor:2001}
	H.~Gildor and E.~Tziperman.
	\newblock {A sea ice climate switch mechanism for the 100-kyr glacial cycles}.
	\newblock {\em Journal of Geophysical Research: Oceans}, 106(C5):9117--9133,
	May 2001.
	
	\bibitem{Gregory:2004io}
	Jonathan~M. Gregory.
	\newblock {A new method for diagnosing radiative forcing and climate
		sensitivity}.
	\newblock {\em Geophysical Research Letters}, 31(3):147--4, 2004.
	
	\bibitem{herein2016}
	M\'aty\'as Herein, J\'anos M\'arfy, G\'abor Dr\'otos, and Tam\'as T\'el.
	\newblock Probabilistic concepts in intermediate-complexity climate models: A
	snapshot attractor picture.
	\newblock {\em Journal of Climate}, 29(1):259--272, 2016.
	
	\bibitem{hogg2008glacial}
	Andrew~McC. Hogg.
	\newblock Glacial cycles and carbon dioxide: A conceptual model.
	\newblock {\em Geophysical research letters}, 35(1), 2008.
	
	\bibitem{Huber2011}
	M~Huber and Rodrigo Caballero.
	\newblock {The early Eocene equable climate problem revisited}.
	\newblock {\em Clim. Past}, 7(2):603--633, 2011.
	
	\bibitem{IPCC2013}
	IPCC.
	\newblock {\em Climate change 2013: the physical science basis: Working Group I
		contribution to the Fifth assessment report of the Intergovernmental Panel on
		Climate Change}.
	\newblock Cambridge University Press, 2013.
	
	\bibitem{Knutti:2008}
	Reto Knutti and Gabriele~C. Hegerl.
	\newblock The equilibrium sensitivity of the earth's temperature to radiation
	changes.
	\newblock {\em Nature Geoscience}, 1:735--743, 2008.
	
	\bibitem{Koehler:2017jj}
	Peter Koehler, Lennert~B. Stap, Anna~S. von~der Heydt, Bas de~Boer,
	Roderik~S.W. Van De~Wal, and J.~Bloch-Johnson.
	\newblock {A State-Dependent Quantification of Climate Sensitivity Based On
		Paleodata of the Last 2.1 Million Years}.
	\newblock {\em Paleoceanography}, 32:1102--1114, 2017.
	
	\bibitem{Koehler:2015}
	P.~K\"ohler, B.~de~Boer, A.S. von~der Heydt, L.B. Stap, and R.S.W. van~de Wal.
	\newblock On the state dependency of the equilibrium climate sensitivity during
	the last 5 million years.
	\newblock {\em Climate of the Past}, 11(12):1801--1823, 2015.
	
	\bibitem{Lenton:2011}
	Timothy~M. Lenton.
	\newblock Early warning of climate tipping points.
	\newblock {\em Nature climate change}, 1(4):201, 2011.
	
	\bibitem{Lenton:2008de}
	Timothy~M. Lenton, Hermann Held, Elmar Kriegler, Jim~W. Hall, Wolfgang Lucht,
	Stefan Rahmstorf, and Hans~Joachim Schellnhuber.
	\newblock {Tipping elements in the Earth's climate system}.
	\newblock {\em Proceedings of the National Academy of Sciences of the United
		States of America}, 105(6):1786--1793, 2008.
	
	\bibitem{lucarini2018}
	Valerio Lucarini.
	\newblock Revising and extending the linear response theory for statistical
	mechanical systems: Evaluating observables as predictors and predictands.
	\newblock {\em Journal of Statistical Physics}, 173:1698--1721, 2018.
	
	\bibitem{Lunt2010ng}
	Daniel~J. Lunt, Alan~M. Haywood, Gavin~A. Schmidt, Ulrich Salzmann, Paul~J.
	Valdes, and Harry~J. Dowsett.
	\newblock {Earth system sensitivity inferred from Pliocene modelling and data}.
	\newblock {\em Nature Geoscience}, 3(1):60--64, 2010.
	
	\bibitem{pfister:2017iz}
	Patrik~L. Pfister and Thomas~F. Stocker.
	\newblock {State-Dependence of the Climate Sensitivity in Earth System Models
		of Intermediate Complexity}.
	\newblock {\em Geophysical Research Letters}, 44(20):10,643--10,653, 2017.
	
	\bibitem{RoeBaker:2007}
	Gerard~H. Roe and Marcia~B. Baker.
	\newblock Why is climate sensitivity so unpredictable?
	\newblock {\em Science}, 318(5850):629--632, 2007.
	
	\bibitem{rohling:2012}
	Eelco~J. Rohling, Appy Sluijs, Henk~A. Dijkstra, Peter K{\"o}hler, Roderick~SW
	van~de Wal, Anna~S von~der Heydt, David~J Beerling, Andr{\'e} Berger, Peter~K
	Bijl, Michel Crucifix, et~al.
	\newblock Making sense of palaeoclimate sensitivity.
	\newblock {\em Nature}, 491(7426):683, 2012.
	
	\bibitem{schwartz:2012}
	Stephen~E Schwartz.
	\newblock Determination of earth's transient and equilibrium climate
	sensitivities from observations over the twentieth century: strong dependence
	on assumed forcing.
	\newblock {\em Surveys in geophysics}, 33(3-4):745--777, 2012.
	
	\bibitem{sellers1969global}
	William~D. Sellers.
	\newblock A global climatic model based on the energy balance of the
	earth-atmosphere system.
	\newblock {\em Journal of Applied Meteorology}, 8(3):392--400, 1969.
	
	\bibitem{Steffen:2018}
	Will Steffen, Johan Rockstr{\"o}m, Katherine Richardson, Timothy~M. Lenton,
	Carl Folke, Diana Liverman, Colin~P. Summerhayes, Anthony~D. Barnosky,
	Sarah~E. Cornell, Michel Crucifix, Jonathan~F. Donges, Ingo Fetzer, Steven~J.
	Lade, Marten Scheffer, Ricarda Winkelmann, and Hans~Joachim Schellnhuber.
	\newblock Trajectories of the earth system in the anthropocene.
	\newblock {\em Proceedings of the National Academy of Sciences},
	115(33):8252--8259, 2018.
	
	\bibitem{Tantet2018}
	Alexis Tantet, Valerio Lucarini, Frank Lunkeit, and Henk~A. Dijkstra.
	\newblock Crisis of the chaotic attractor of a climate model: a transfer
	operator approach.
	\newblock {\em Nonlinearity}, 31(5):2221--2251, apr 2018.
	
	\bibitem{Tantet2015}
	Alexis Tantet, Fiona~R. van~der Burgt, and Henk~A. Dijkstra.
	\newblock An early warning indicator for atmospheric blocking events using
	transfer operators.
	\newblock {\em Chaos: An Interdisciplinary Journal of Nonlinear Science},
	25(3):036406, 2015.
	
	\bibitem{Heydt:2017}
	Anna~S. von~der Heydt and Peter Ashwin.
	\newblock {State dependence of climate sensitivity: attractor constraints and
		palaeoclimate regimes}.
	\newblock {\em Dynamics and Statistics of the Climate System}, 1(1):1--21,
	February 2017.
	
	\bibitem{Heydt2016rev}
	Anna~S. von~der Heydt, Henk~A. Dijkstra, Roderik~S.W. Van De~Wal, Rodrigo
	Caballero, Michel Crucifix, Gavin~L. Foster, M.~Huber, Peter Koehler, E.~J.
	Rohling, Paul~J. Valdes, Peter Ashwin, Sebastian Bathiany, T.~Berends, L.~van
	Bree, Peter~D. Ditlevsen, Michael Ghil, Alan~M. Haywood, Joel Katzav, Gerrit
	Lohmann, J.~Lohmann, Valerio Lucarini, A.~Marzocchi, H.~P\"{a}like,
	I.~Ruvalcaba~Baroni, D~Simon, Appy Sluijs, L.B. Stap, A.~Tantet, J.P.
	Viehbahn, and Martin Ziegler.
	\newblock {Lessons on Climate Sensitivity From Past Climate Changes}.
	\newblock {\em Current Climate Change Reports}, 2(4):148--158, 2016.
	
	\bibitem{Heydt2014}
	Anna~S. von~der Heydt, Peter Koehler, Roderik~S.W. Van De~Wal, and Henk~A.
	Dijkstra.
	\newblock {On the state dependency of fast feedback processes in (paleo)
		climate sensitivity}.
	\newblock {\em Geophysical Research Letters}, 41(18):6484--6492, 2014.
	
	\bibitem{zaliapin2010another}
	I.~Zaliapin and M.~Ghil.
	\newblock Another look at climate sensitivity.
	\newblock {\em Nonlinear Processes in Geophysics}, 17(2):113--122, 2010.
	
	\bibitem{Zhang:2015}
	Xiaozhu Zhang, Christian Kuehn, and Sarah Hallerberg.
	\newblock Predictability of critical transitions.
	\newblock {\em Phys. Rev. E}, 92:052905, Nov 2015.
	
\end{thebibliography}

\newpage

\appendix

\section{Appendix: Tipping in a multi-box climate model with ocean biogeochemistry}
\label{sec:GT}

To investigate the notion of two-point sensitivity in a more complex and physics based earth system model, we explore here the  Gildor-Tziperman (2001) model \cite{Gildor:2001} with Milaknovich forcing and biogeochemistry in the ocean. In this model the ocean and atmosphere consists of 4 latitudinal (zonally averaged) boxes each, with two layers in the ocean and one in the atmosphere. Within the boxes a variety of thermodynamic quantities (eg temperature $T$) and species (eg ocean salinity $S$ and atmospheric \coo{}) are modelled. In addition there is dynamic land ice (slowly evolving) and sea ice (rapidly evolving) as well as fluxes that join these boxes. The system is sufficiently complex to allow modelling of the Pleistocene ice-age oscillations of land-ice in response to Milankovich forcing. The glacial-interglacial cycles appear in this model as internally generated self-sustained oscillations, which are then modified by the Milankovich forcing. More details are given in \cite{Gildor:2001,Heydt:2017}.

\begin{figure}[!ht]
\centering
\includegraphics[width=8cm]{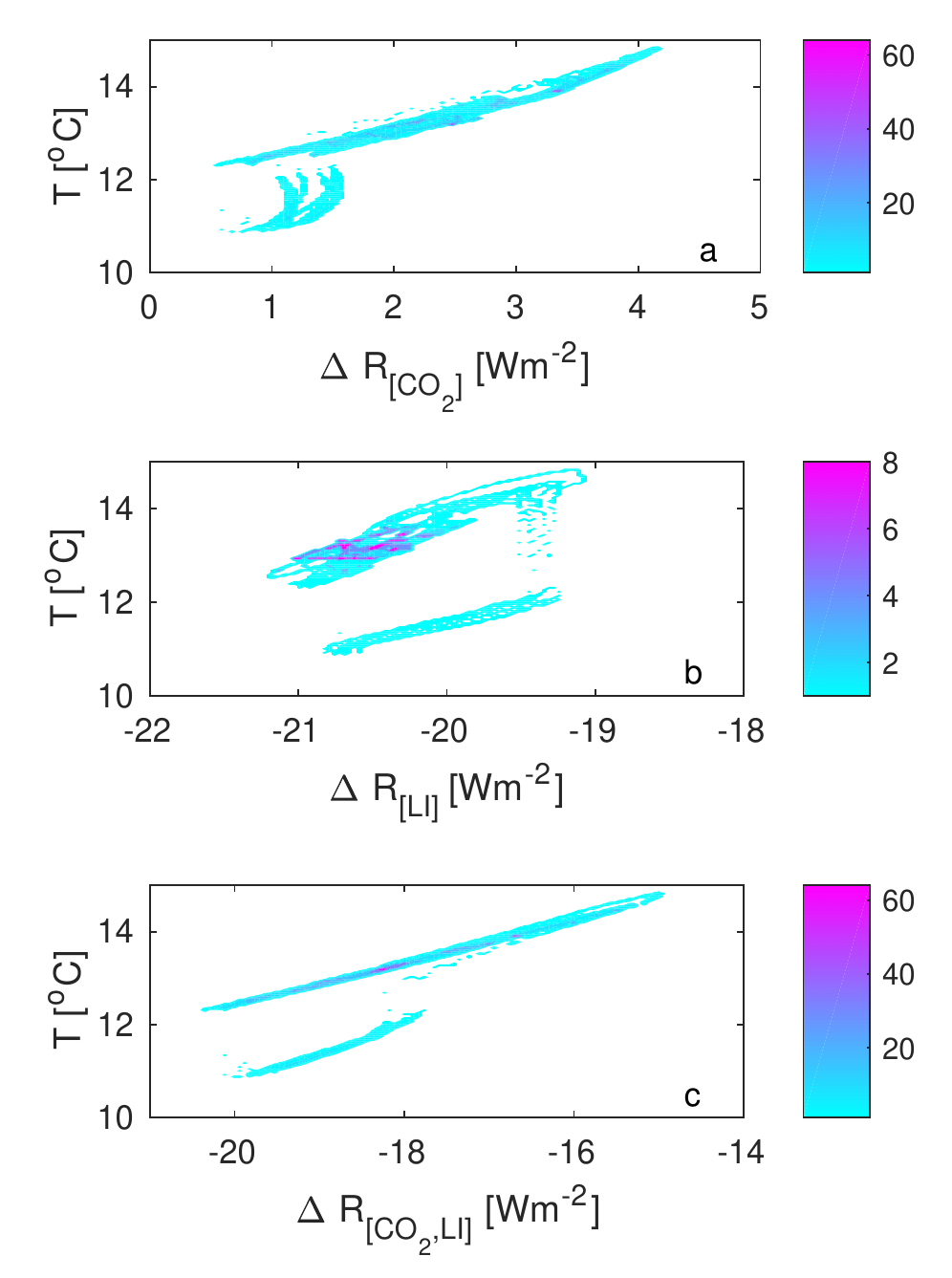}
\caption{Plots of $T$ vs $\Delta R$ for the Gildor-Tziperman model \cite{Gildor:2001}; see \cite{Heydt:2017}. (a) projection onto the $(\Delta R_{[CO_2]},T)$-plane, which should reflect the Earth System Sensitivity only considering \coo{} as forcing; (b) projection onto the $(\Delta R_{[LI]},T)$-plane, considering only the slow land-ice albedo feedback as forcing; (c) projection onto the $(\Delta R_{[CO_2,LI]},T)$-plane, considering both \coo{} and the slow feedbacks (in this model there is only the land-ice albedo feedback slow) as forcing. Following the formalism of \cite{rohling:2012} the slopes in this graph should reflect a good approximation of the Charney ECS.}
\label{fig:GT_attractor}
\end{figure}

Figure~\ref{fig:GT_attractor}
shows projections of a long trajectory (500 kyr) on the climate attractor of this more complex climate model onto the plane of global mean temperature $T$ against (a) $\Delta R_{[CO_2]}$ (b) $\Delta R_{[LI]}$ and (c) $\Delta R_{[CO_2,LI]}$ \cite{Heydt:2017}. Observe that all three projections clearly show two climate regimes, a lower `cold' state (corresponding to large amounts of sea ice and $T<12.28 C$) and an upper `warm' state (corresponding to $T>12.28 C$). The projection on the combined radiative forcing of \coo{} and the slow feedback in land ice changes $\Delta R_{[LI]}$ shows a clear slope and hence 'mean' sensitivities in both regimes (Figure ~\ref{fig:GT_attractor}c). Following the formalism of \cite{rohling:2012}, the slopes in (a) should reflect the Earth System sensitivity, while (c) should give a good approximation of the Charney or equilibrium climate sensitivity. Note that in this model there is only one slow feedback to correct for, namely the land-ice albedo feedback. 
When projecting onto the $\Delta R_{[CO_2]}$ plane (Figure ~\ref{fig:GT_attractor}a) the cold regime appears very diffuse and with very high (or sometimes negative) Earth System sensitivities, suggesting that in the cold branch local climate dynamics are not entirely determined by \coo. Similarly, when projecting only on land ice changes $\Delta R_{{LI}}$ (Figure ~\ref{fig:GT_attractor}b) the warm branch appears more diffuse, i.e. \coo{} dynamics seem to be more important than land ice dynamics on this branch.

\begin{figure}[!ht]
\centering
\includegraphics[height=9cm]{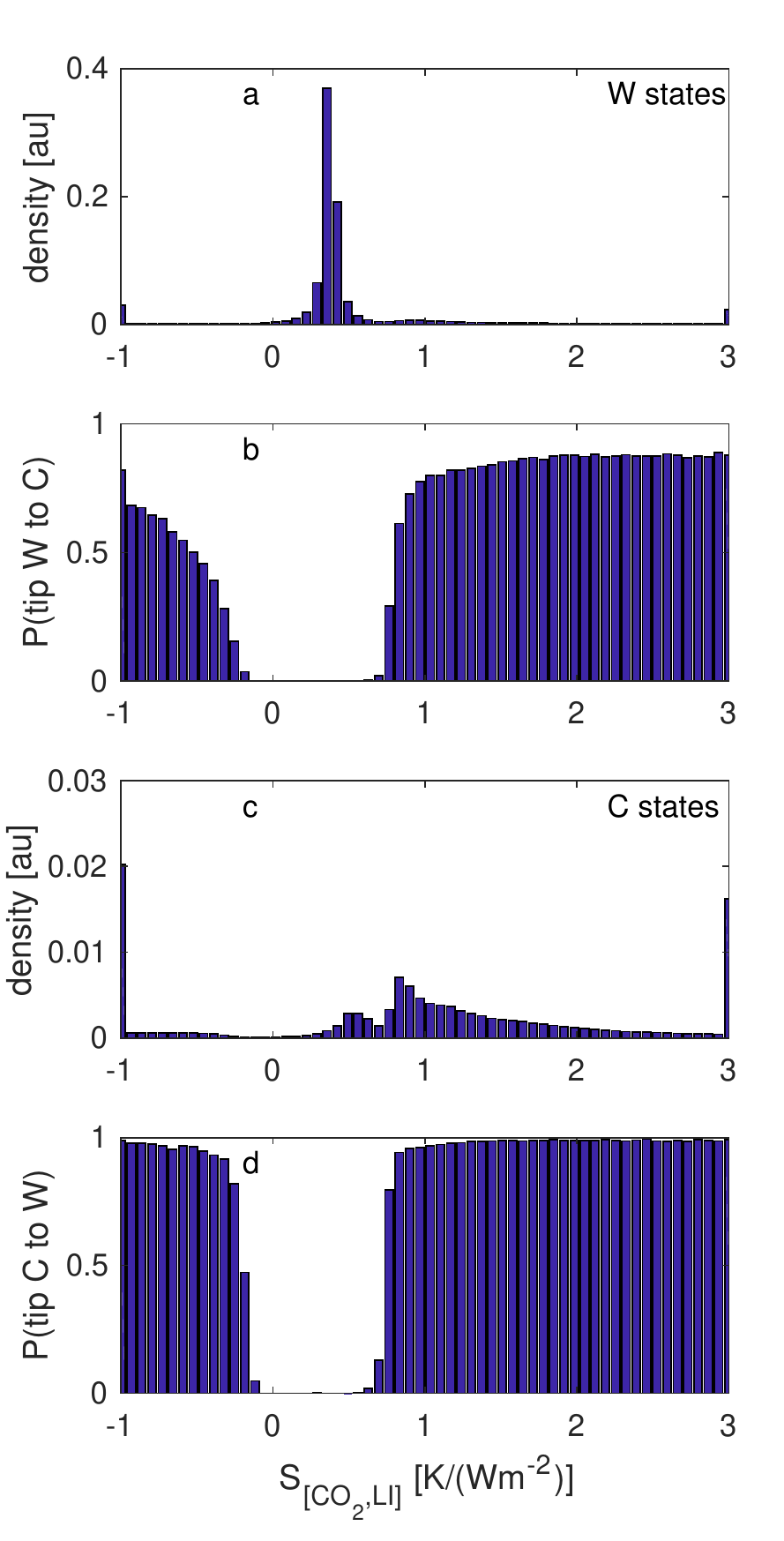}\includegraphics[height=9cm]{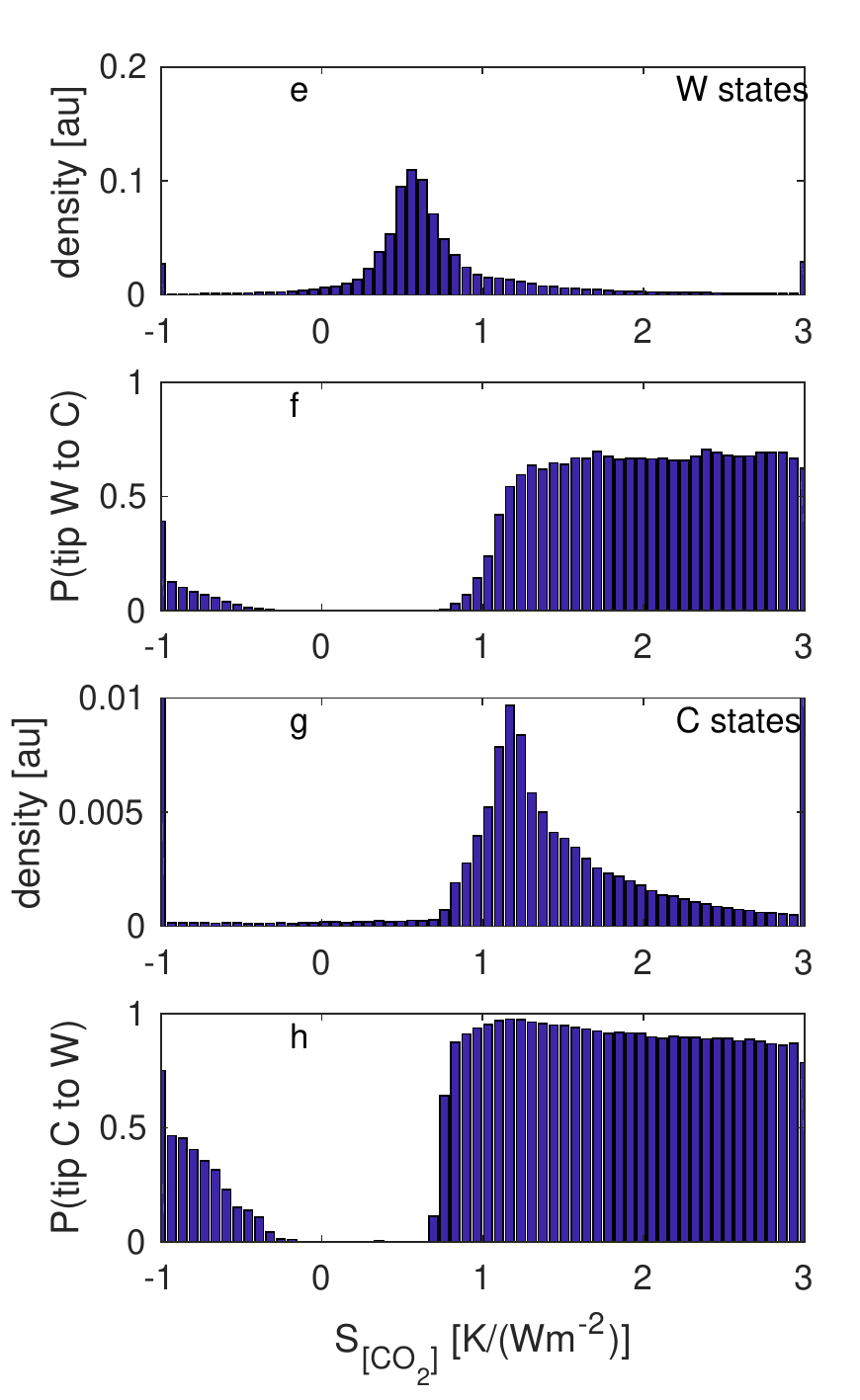}
\caption{Conditional sensitivities and probabilities of tipping for glacial cycles simulations with the Gildor-Tziperman model \cite{Gildor:2001} (as in Figure~\ref{fig:EBMRTsensitivities} for the energy balance model). (a-d)  Conditional two-point sensitivities compensating for slow feedbacks (i.e. reflecting ECS) and probabilities of tipping (i.e. from the distribution shown in Figure~\ref{fig:GT_attractor}c). (e-h) Conditional earth system sensitivities ESS (not compensated for slow feedbacks) and probabilities of tipping (i.e. from the distribution shown in Figure~\ref{fig:GT_attractor}a).}
\label{fig:GTsensitivities}
\end{figure}

The left column of Figure~\ref{fig:GTsensitivities} (a-d) shows the distribution of two-point climate sensitivities for $R_{[CO_2,LI]}$ conditional on regime: this is comparable to Figure~\ref{fig:EBMRTsensitivities}(a-d) in that one can observe (i) clearly localised distributions of ECS in (a,c) conditional on remaining within the W or C regime, (ii) a broader distribution in (c): this seems to be associated with the curvature of the C regime branch in Figure~\ref{fig:GT_attractor}c), (iii) a clear association of tipping from W to C (b) or from C to W (d) being associated with extreme sensitivities. Note that for this model there is no energy balance model available and so it is not possible to compute the instantaneous sensitivity.
For comparison we show in the right column of Figure~\ref{fig:GTsensitivities} (e-h) the same distributions for the Earth System Sensitivities (ESS), which are not compensated for the slow feedback (in this model the land-ice albedo feedback). Observe that for both regimes there is a much broader distribution for the ESS (e and g) than for the ECS in (a) and (c). In particular in the cold regime, earth system sensitivities (g) are much higher than equilibrium sensitivities (c) because the land-ice albedo feedback is very strong in the cold (land-ice covered) states. 

\end{document}